\documentclass[%
reprint,
superscriptaddress,
nofootinbib,
 amsmath,amssymb,
 aps,
]{revtex4-1}
\usepackage{graphicx}
\usepackage{subfigure}
\usepackage{dcolumn}
\usepackage{bm}
\usepackage[%
bookmarksnumbered,
bookmarksopen,
colorlinks,
citecolor=blue,
linkcolor=blue,
]{hyperref} 

\def\esym{$E_{sym}(\rho)$~}

\def\es0{$E_{sym}(\rho_0)$~}
\begin{document}


\title{Effects of the momentum dependence of nuclear symmetry potential on pion observables in Sn + Sn collisions at 270 MeV/nucleon}

\author{Gao-Feng Wei}\email[ E-mail: ]{wei.gaofeng@gznu.edu.cn}
\affiliation{School of Physics and Electronic Science, Guizhou Normal University, Guiyang 550025, China}
\affiliation{Guizhou Provincial Key Laboratory of Radio Astronomy and Data Processing, Guizhou Normal University, Guiyang 550025, China}
\author{Xin Huang}
\affiliation{School of Physics and Electronic Science, Guizhou Normal University, Guiyang 550025, China}
\author{Qi-Jun Zhi}
\affiliation{School of Physics and Electronic Science, Guizhou Normal University, Guiyang 550025, China}
\affiliation{Guizhou Provincial Key Laboratory of Radio Astronomy and Data Processing, Guizhou Normal University, Guiyang 550025, China}
\author{Ai-Jun Dong}
\affiliation{School of Physics and Electronic Science, Guizhou Normal University, Guiyang 550025, China}
\affiliation{Guizhou Provincial Key Laboratory of Radio Astronomy and Data Processing, Guizhou Normal University, Guiyang 550025, China}

\author{Chang-Gen Peng}
\affiliation{Guizhou Provincial Key Laboratory of Public Big Data, Guizhou University, Guiyang 550025, China}
\author{Zheng-Wen Long}
\affiliation{College of Physics, Guizhou University, Guiyang 550025, China}


\begin{abstract}
	
Within a transport model, we study effects of the momentum dependence of nuclear symmetry potential on pion observables in central Sn + Sn collisions at 270 MeV/nucleon. To this end, a quantity $U_{sym}^{\infty}(\rho_{0})$, i.e., the value of nuclear symmetry potential at the saturation density $\rho_{0}$ and infinitely large nucleon momentum, is used to characterise the momentum dependence of nuclear symmetry potential. It is shown that with a certain $L$ (i.e., slope of nuclear symmetry energy at $\rho_{0}$) the characteristic parameter $U_{sym}^{\infty}(\rho_{0})$ of symmetry potential affects significantly the production of $\pi^{-}$ and $\pi^{+}$ as well as their pion ratios. Moreover, through comparing the charged pion yields, pion ratios as well the spectral pion ratios of theoretical simulations for the reactions $^{108}$Sn + $^{112}$Sn and $^{132}$Sn + $^{124}$Sn with the corresponding data in S$\pi$RIT experiments, 
we find that our results favor a constraint on $U_{sym}^{\infty}(\rho_{0})$, i.e., $-160^{+18}_{-9}$~MeV, and the $L$ is also suggested within a range, i.e., $62.7<L<93.1$~MeV.
In addition, it is shown that the pion observable of $^{197}$Au + $^{197}$Au collisions at 400~MeV/nucleon also supports the extracted value for $U_{sym}^{\infty}(\rho_{0})$. 

\end{abstract}

\maketitle


\section{introduction}\label{introduction}

The equation of state (EoS) of asymmetric nuclear matter (ANM) especially its nuclear symmetry energy \esym term plays an essential role in studying the structure and evolution of radioactive nuclei as well as the synthesis of medium and heavy nuclei~\cite{Typel01,kolo05,ditoro,LCK08,Tam11,Vin14,Hor14,Baldo16,Rein16,MCW18,Yu20,MCW21}. The \esym characterizes the variation of EoS of the symmetric nuclear matter (SNM) to that of the pure neutron matter (PNM), the latter is closely connected to the neutron star (NS) matter. Naturally, the properties of NS such as the radius as well as the deformation of NS merger are also closely related to the \esym especially that at densities of about twice the saturation density $\rho_{0}$~\cite{Estee21,Tsang19,Lim18,Tews18,Drago14,Steiner12,Duco11,Latt16}.
Nevertheless, knowledge on the \esym at suprasaturation densities is still far from satisfactory so far, although that around and below $\rho_{0}$~\cite{Brown13,LiBA16} as well as the isospin-independent part of EoS for ANM, i.e., EoS of SNM~\cite{Dan02,Oert17,Cai17}, are relatively well determined. 
Essentially, the EoS of ANM and its \esym term are determined by the nuclear mean field especially its isovector part, i.e., the symmetry/isovector potential~\cite{Liu21,Wei20a}. However, because of the extreme challenge of relatively direct detection of isovector potential in experiments, one only extracted using the nucleon-nucleus scattering and ($p$,$n$) charge-exchange reactions between isobaric analog states limited information of isovector potential at $\rho_{0}$, and parameterized as $U_{sym}(\rho_{0},E_{k})=a - bE_{k}$, where $a~{\approx}~22 - 34$ MeV, $b~{\approx}~0.1 - 0.2$ and $E_{k}$ is limited to no more than 200 MeV~\cite{Hoff72,Kon03,Jeu91}. 

Heavy-ion collision (HIC) is one of the most promising approaches to explore the symmetry potential/energy especially at suparsaturation densities~\cite{Estee21,Jhang21,Shane15,ditoro,LCK08,FOPI}. 
Very recently, the S$\pi$RIT collaboration reported the results from the first measurement dedicated to probe the \esym at suprasaturation densities via pion production in Sn + Sn collisions at 270 MeV/nucleon carried out at RIKEN-RIBF in Japan~\cite{Jhang21}. 
Moreover, they compared the charged pion yields as well as their single and double pion ratios with the corresponding simulation results from seven transport models. 
Qualitatively, the theoretical simulations from seven transport models reach an agreement with the data, yet quantitatively, almost all the models cannot very satisfactorily reproduce both the pion yields and their single as well as double pion ratios of the experimental data~\cite{Jhang21}. To this situation, author of Ref.~\cite{Yong21} claimed that through considering about 20\% high momentum nucleons 
in colliding nuclei can reproduce quite well both the charged pion yields and their pion ratios of the experimental data, due to the 
high momentum distribution in nuclei caused by the short-range correlations (SRCs)
~\cite{Subedi08,Wein11,Sar14,Ciofi15,Ohen14,Ohen18}. Following this work, we focus on the momentum dependence of symmetry potential since that plays a more important role in probing the high density behavior of \esym\cite{Brue64,Dabr73,Gior10}.
Actually, as indicated in Ref.~\cite{Jhang21} as well in a series of literatures~\cite{trans1,trans2,trans3,trans4,trans5} of transport model comparison project, the possible reasons for the unsatisfactory of seven models quantitatively fitting experimental data may be different assumptions regarding the mean field potential, pion potential as well as the treatment of Coulomb field. Therefore, it is very necessary to explore how the momentum dependence of symmetry potential affects the pion production in HICs. As to other factors mentioned above, we also give detailed consideration according to some sophisticated treatment ways
as discussed in Sec.~\ref{Model}. In Sec.~\ref{Results and Discussions}, we discuss the results of the present study. A summary is given finally in Sec.~\ref{Summary}.

\section{The Model}\label{Model}

This study is carried out within an isospin- and momentum-dependent Boltzmann-Uehling-Uhlenbeck (IBUU) transport model. In the framework, the present model originates from the IBUU04~\cite{Das03,IBUU} and/or IBUU11~\cite{CLnote} models. However, the present model has been greatly improved to more accurately simulate pion production as discussed in the following.

First, a separate density-dependent scenario for the in-medium nucleon-nucleon interaction~\cite{Xu10,Chen14,Wei20}, i.e.,
\begin{equation}\label{den-term}
v_{\small D}=t_{0}(1+x_{0}P_{\sigma})[\rho_{\tau_{i}}(\textbf{r}_{i}){\small +}\rho_{\tau_{j}}(\textbf{r}_{j})]^{\alpha}\delta(\textbf{r}_{ij}),
\end{equation}
is used to replace the density-dependent term of original Gogny effective interaction~\cite{Gogny80}, i.e.,
\begin{eqnarray}\label{Gongy}
v(r)&=&\sum_{i=1,2}(W+BP_{\sigma}-HP_{\tau}-MP_{\sigma}P_{\tau})_{i}e^{-r^{2}/\mu_{i}^{2}}\notag \\
&+&t_{0}(1+x_{0}P_{\sigma}){\big[}\rho{\big(}\frac{{\small\textbf{r}_{i}}+{\small\textbf{r}_{j}}}{{\small{2}}}{\big)}{\big]}^{\alpha}\delta(\textbf{r}_{ij}),
\end{eqnarray}
where $W$, $B$, $H$, $M$, and $\mu$ are five parameters, $P_{\tau}$ and $P_{\sigma}$ are the isospin and spin exchange operators, respectively; while $\alpha$ is the density dependent parameter used to mimic in-medium effects of the many-body interactions~\cite{Xu10,Chen14,Wei20}. As indicated in Ref.~\cite{Duguet03}, the separate density dependence of effective two-body interactions is originated from the renormalization of multibody force effects, and the latter may extend the density dependence of effective interactions for calculations beyond the mean-field approximation. Moreover, the nuclear structure studies have already shown that, with the separate density-dependent scenario for the in-medium nucleon-nucleon interaction, the more satisfactory results, e.g., the binding energies, single-particle energies, and electron scattering cross sections for $^{16}$O, $^{40}$Ca, $^{48}$Ca, $^{90}$Zr and $^{208}$Pr~\cite{Negele70}, can be reached compared with the corresponding experiments.
Correspondingly, the potential energy density for ANM with this improved momentum-dependent interaction (IMDI) is expressed~\cite{Chen14} as
\begin{eqnarray}
V(\rho,\delta) &=&\frac{A_{u}(x)\rho_{n}\rho_{p}}{\rho_{0}}%
+\frac{A_{l}(x)}{2\rho_{0}}(\rho_{n}^{2}+\rho_{p}^{2})+\frac{B}{\sigma + 1}\frac{\rho^{\sigma+1}}{\rho_{0}^{\sigma}}  \notag \\
&\times&\Big{\{}\frac{1+x}{2}(1-\delta^{2})+\frac{1-x}{4}\big{[}(1+\delta)^{\sigma+1}+(1-\delta)^{\sigma+1}\big{]}\Big{\}}
\notag \\
&+&\frac{1}{\rho _{0}}\sum_{\tau,\tau^{\prime}}C_{\tau,\tau^{\prime}}\int\int d^{3}pd^{3}p^{\prime }\frac{f_{\tau }(%
	\vec{r},\vec{p})f_{\tau^{\prime}}(\vec{r},\vec{p}^{\prime})}{1+(\vec{p}-\vec{p}^{\prime })^{2}/\Lambda ^{2}}.
\label{IMDI}
\end{eqnarray}%
In the mean-field approximation, Eq.~(\ref{IMDI}) leads to the following single-nucleon potential for the present model~\cite{Xu10,Chen14,Wei20},
\begin{eqnarray}
U(\rho,\delta ,\vec{p},\tau ) &=&A_{u}(x)\frac{\rho _{-\tau }}{\rho _{0}}%
+A_{l}(x)\frac{\rho _{\tau }}{\rho _{0}}+\frac{B}{2}{\big(}\frac{2\rho_{\tau} }{\rho _{0}}{\big)}^{\sigma }(1-x)  \notag \\
&+&\frac{2B}{%
	\sigma +1}{\big(}\frac{\rho}{\rho _{0}}{\big)}^{\sigma }(1+x)\frac{\rho_{-\tau}}{\rho}{\big[}1+(\sigma-1)\frac{\rho_{\tau}}{\rho}{\big]}
\notag \\
&+&\frac{2C_{l }}{\rho _{0}}\int d^{3}p^{\prime }\frac{f_{\tau }(%
	\vec{p}^{\prime })}{1+(\vec{p}-\vec{p}^{\prime })^{2}/\Lambda ^{2}}
\notag \\
&+&\frac{2C_{u }}{\rho _{0}}\int d^{3}p^{\prime }\frac{f_{-\tau }(%
	\vec{p}^{\prime })}{1+(\vec{p}-\vec{p}^{\prime })^{2}/\Lambda ^{2}},
\label{IMDIU}
\end{eqnarray}%
where $\sigma=\alpha+1$, $\tau=1$ for neutrons and $-1$ for protons, and the parameters $A_{u}(x)$, $A_{l}(x)$, $C_{u}(\equiv C_{\tau,-\tau})$ and $C_{l}(\equiv C_{\tau,\tau})$ are expressed as
\begin{eqnarray}
A_{l}(x)&=&A_{l0} + U_{sym}^{\infty}(\rho_{0}) - \frac{2B}{\sigma+1}\notag \label{formulaAl}\\
&\times&\Big{[}\frac{(1-x)}{4}\sigma(\sigma+1)-\frac{1+x}{2}\Big{]},  \\
A_{u}(x)&=&A_{u0} - U_{sym}^{\infty}(\rho_{0}) + \frac{2B}{\sigma+1}\notag \label{formulaAu}\\
&\times&\Big{[}\frac{(1-x)}{4}\sigma(\sigma+1)-\frac{1+x}{2}\Big{]},\\
C_{l}&=&C_{l0} - 2U_{sym}^{\infty}(\rho_{0})\frac{p_{f0}^{2}}{\Lambda^{2}\ln \big{[}(4p_{f0}^{2}+\Lambda^{2})/\Lambda^{2}\big{]}},\label{formulaCl}\\
C_{u}&=&C_{u0} + 2U_{sym}^{\infty}(\rho_{0})\frac{p_{f0}^{2}}{\Lambda^{2}\ln \big{[}(4p_{f0}^{2}+\Lambda^{2})/\Lambda^{2}\big{]}},
\label{formulaCu}
\end{eqnarray}
where $p_{f0}$ is the nucleon Fermi momentum in SNM at $\rho_{0}$, and $U_{sym}^{\infty}(\rho_{0})$ proposed in Ref.~\cite{CLnote} is used to characterize the momentum dependence of symmetry potential at $\rho_{0}$. As to the derivation of Eqs.~(\ref{formulaAl})-(\ref{formulaCu}) as well the expression of symmetry potential and/or energy, see Ref.~\cite{Chen14} for the details.

\begin{figure}[t]
	\includegraphics[width=0.86\columnwidth]{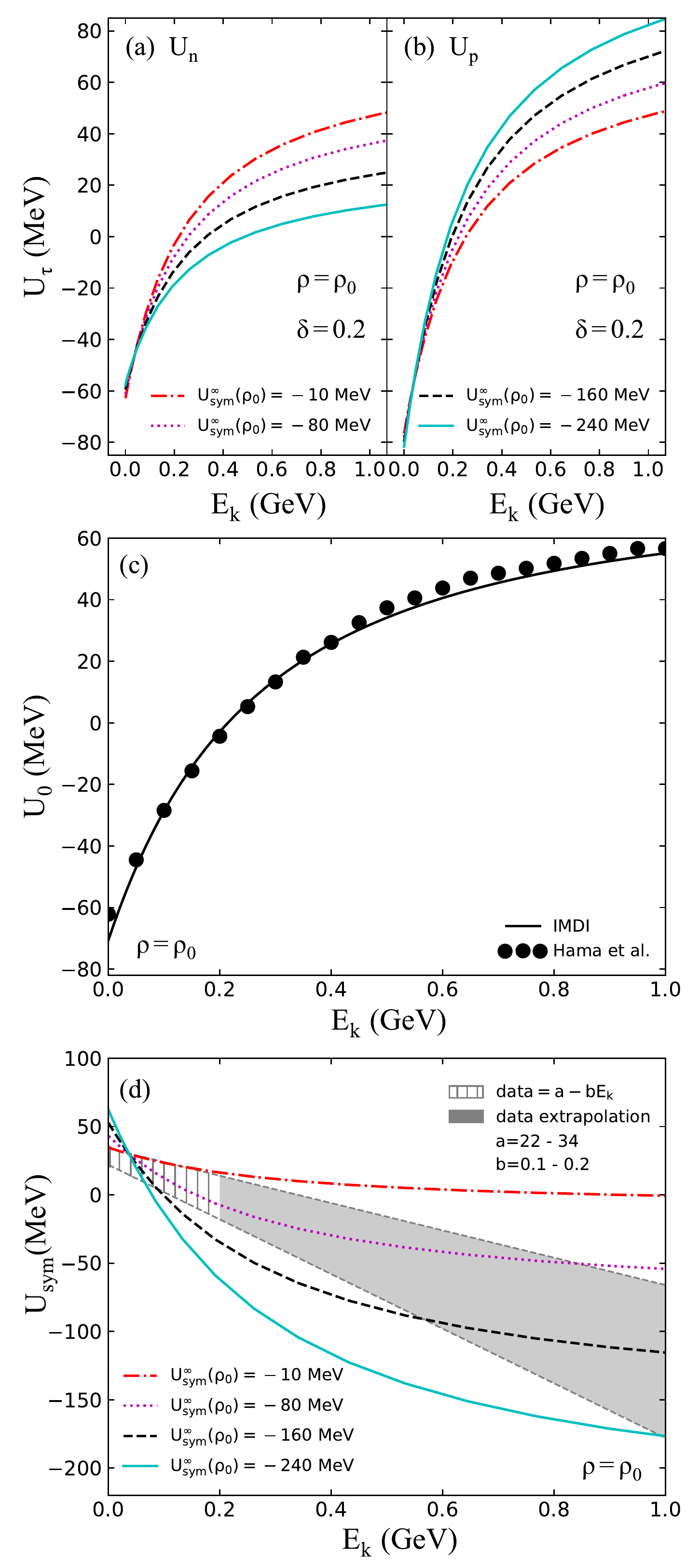}
	\caption{(Color online) Kinetic-energy dependent neutron (a) and proton (b) potentials as well the isoscalar (c) and isovector (d) potentials at $\rho_{0}$ calculated from the IMDI interaction. The Schr\"{o}dinger-equivalent isoscalar potential obtained by Hama {\it et al}. and the parameterized isovector potential from the experimental and/or empirical data are also shown to compare with the isoscalar and isovector potentials calculated from the IMDI interaction.}
	\label{pot}
\end{figure}

Presently, knowledge on the momentum dependence of symmetry potential even at $\rho_{0}$ is rather limited as aforementioned~\cite{Hoff72,Kon03,Jeu91}. Therefore, taking the parameterized symmetry potential as a reference, we treat the $U_{sym}^{\infty}(\rho_{0})$ as a free parameter similar as the $x$ parameter, which is used to mimic the slope value $L\equiv{3\rho({dE_{sym}}/d\rho})$ of \esym at $\rho_{0}$ without changing the value of $E_{sym}(\rho)$ at $\rho_{0}$ and any properties of the SNM. Actually, a similar quantity (i.e., $y$ parameter) in Refs.~\cite{Xu15,Xu15b,Xu17} has been used to describe the momentum dependence of symmetry potential at $\rho_{0}$, however, the quantitative constraints on it are not concluded. In addition, it should be mentioned that the $B$-terms in Eqs.~(\ref{IMDI}) and (\ref{IMDIU}) as well as in the expressions of $A_{u}$ and $A_{l}$ are different from those in Refs.~\cite{Xu15,Xu15b,Xu17}. This is exactly because the separate density-dependent scenario for in-medium nucleon-nucleon interaction has been adopted in the present model to more delicate treatment of the in-medium many-body force effects~\cite{Chen14}, that also affects significantly the pion production in HICs~\cite{Wei20}. The seven parameters $A_{l0}$, $A_{u0}$, $B$, $\sigma$, $C_{l0}$, $C_{u0}$ and $\Lambda$ are determined by fitting seven experimental and/or empirical constraints on properties of nuclear matter at $\rho_{0}=0.16$~fm$^{-3}$. The first six quantities are the binding energy $-16$~MeV, the pressure $P_{0}=0$~MeV/fm$^{3}$, the incompressibility $K_{0}=230$~MeV for SNM, the isoscalar effective mass $m^{*}_{s}=0.7m$, the isoscalar potential at infinitely large nucleon momentum $U^{\infty}_{0}(\rho_{0})=75$~MeV, as well as the symmetry energy $E_{sym}(\rho_{0})=32.5$~MeV, and the seventh quantity is the considered $U_{sym}^{\infty}(\rho_{0})$. The values of these parameters are $A_{l0}=A_{u0}=-66.963$~MeV, $B=141.963$~MeV, $C_{l0}=-60.486$~MeV, $C_{u0}=-99.702$~MeV, $\sigma=1.2652$, and $\Lambda=2.424p_{f0}$. 

\begin{figure}[t]
	\includegraphics[width=0.9\columnwidth]{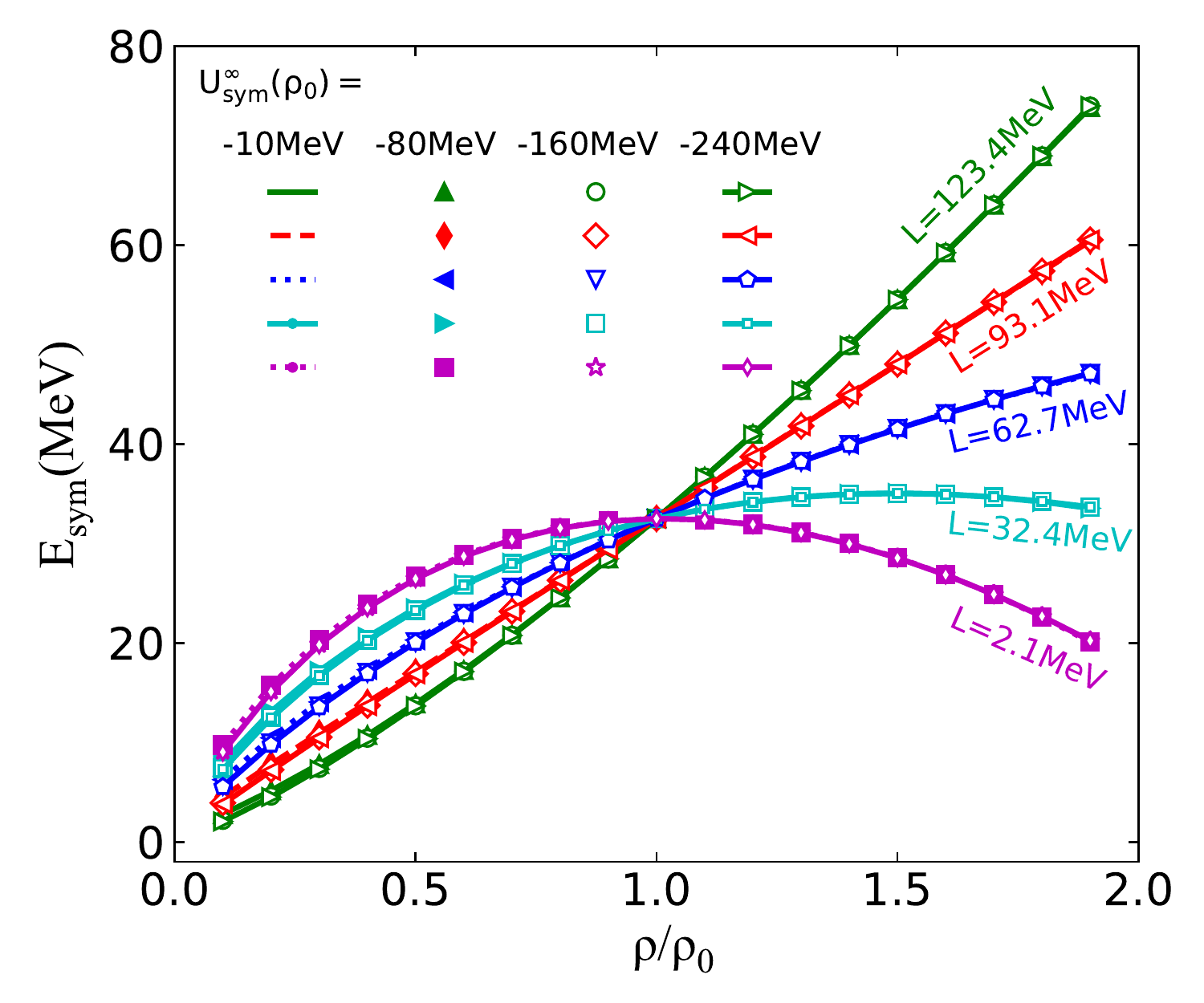}
	\caption{(Color online) Density dependence of the \esym with different $U_{sym}^{\infty} (\rho_{0})$ calculated from the IMDI interaction.} \label{esym}
\end{figure}

Shown in upper windows of Fig.~\ref{pot} are the kinetic-energy dependent neutron and proton potentials at $\rho_{0}$ with different $U_{sym}^{\infty} (\rho_{0})$ calculated from the IMDI interaction. 
It is seen that, as the increase of $|U_{sym}^{\infty} (\rho_{0})|$, the neutron potential shows the opposite variation tendency compared with the proton potential due to the isospin effects. Shown in middle and lower windows of Fig.~\ref{pot} are the isoscalar and isovector potentials at $\rho_{0}$ in comparison with, respectively, the Schr\"{o}dinger-equivalent isoscalar potential obtained by Hama {\it et al}.~\cite{Hama90,Buss12} and the parameterized isovector potential from the experimental and/or empirical data~\cite{Hoff72,Kon03,Jeu91}. To provide more intuitive references for $U_{sym}^{\infty}(\rho_{0})$, we also extrapolate the experimental and/or empirical isovector potential to nucleon kinetic energy up to 1 GeV. Obviously, quite good consistency can be seen for the isoscalar potential between the present model and that of the Hama {\it et al}.  
Moreover, the values of our symmetry potentials at the Fermi kinetic energy (i.e., about 36.8 MeV) even with different $U_{sym}^{\infty} (\rho_{0})$ are the same and also within the allowed range of experimental and/or empirical data. Actually, it is exactly based on the values of symmetry potentials at the Fermi kinetic energy and the infinitely nucleon momentum that we determine the momentum dependence of symmetry potential at $\rho_{0}$. 
On the other hand, since the isoscalar potentials are unchanged with different $U_{sym}^{\infty}(\rho_{0})$, one naturally expects the differences of momentum dependence between symmetry potentials with different $U_{sym}^{\infty}(\rho_{0})$ can be reflected by the pion observable in HICs, because the different symmetry potentials can lead to the different isospin effects and thus different $\pi^{-}/\pi^{+}$ ratios for neutron-rich reactions. Therefore, to get the pion observable more cleanly reflecting effects of the momentum dependence of symmetry potentials, 
it is useful to map the momentum dependent symmetry potentials with different $U_{sym}^{\infty}(\rho_{0})$ into cases with the same $E_{sym}(\rho)$. This is carried out by fitting the identical constraints for SNM  as well as the identical slope parameter $L$ of \esym at $\rho_{0}$, the corresponding results are also shown in Fig.~\ref{esym}. It is seen that even with the same $E_{sym}(\rho)$, the corresponding symmetry potential could be very different since the fact that the symmetry potentials depend not only on the nucleon density but also on the nucleon momentum or energy. 

Second, to more accurately simulate pion production in HICs, we also consider the pion potential effects in HICs. 
Specifically, when the pionic momentum is higher than 140 MeV/$c$, we use the pion potential based on the $\Delta$-hole model, of the form adopted in Ref.~\cite{Buss12}; when the pionic momentum is lower than 80 MeV/$c$, we adopt the pion potential of the form used in Refs.~\cite{Eric66,Oset88,Oset93}; while for the pionic momentum falling into the range from 80 to 140 MeV/$c$, an interpolative pion potential constructed in Ref.~\cite{Buss12} is used. The present pion potential includes the isospin- and momentum-dependent pion $s$-wave and $p$-wave potentials in nuclear medium as that in Ref.~\cite{Zhang17}, see  Refs.~\cite{Buss12,Eric66,Oset88,Oset93} for the details. 
The in-medium isospin-dependent baryon-baryon elastic and inelastic scattering cross sections $\sigma_{medium}$ are determined by the corresponding free-space ones $\sigma_{free}$ multiplying by a factor $R_{medium}$, i.e.,
\begin{equation}
\sigma_{medium}=\sigma_{free}R_{medium}(\rho,\delta,\vec{p}),
\end{equation}
where the reduced factor is determined as $R_{medium}=(\mu^{*}_{BB}/\mu_{BB})^{2}$, the $\mu_{BB}$ and $\mu^{*}_{BB}$ are the reduced masses of colliding baryon paris in free-spcace and nuclear medium, respectively.

\begin{figure}[t]
	\includegraphics[width=\columnwidth]{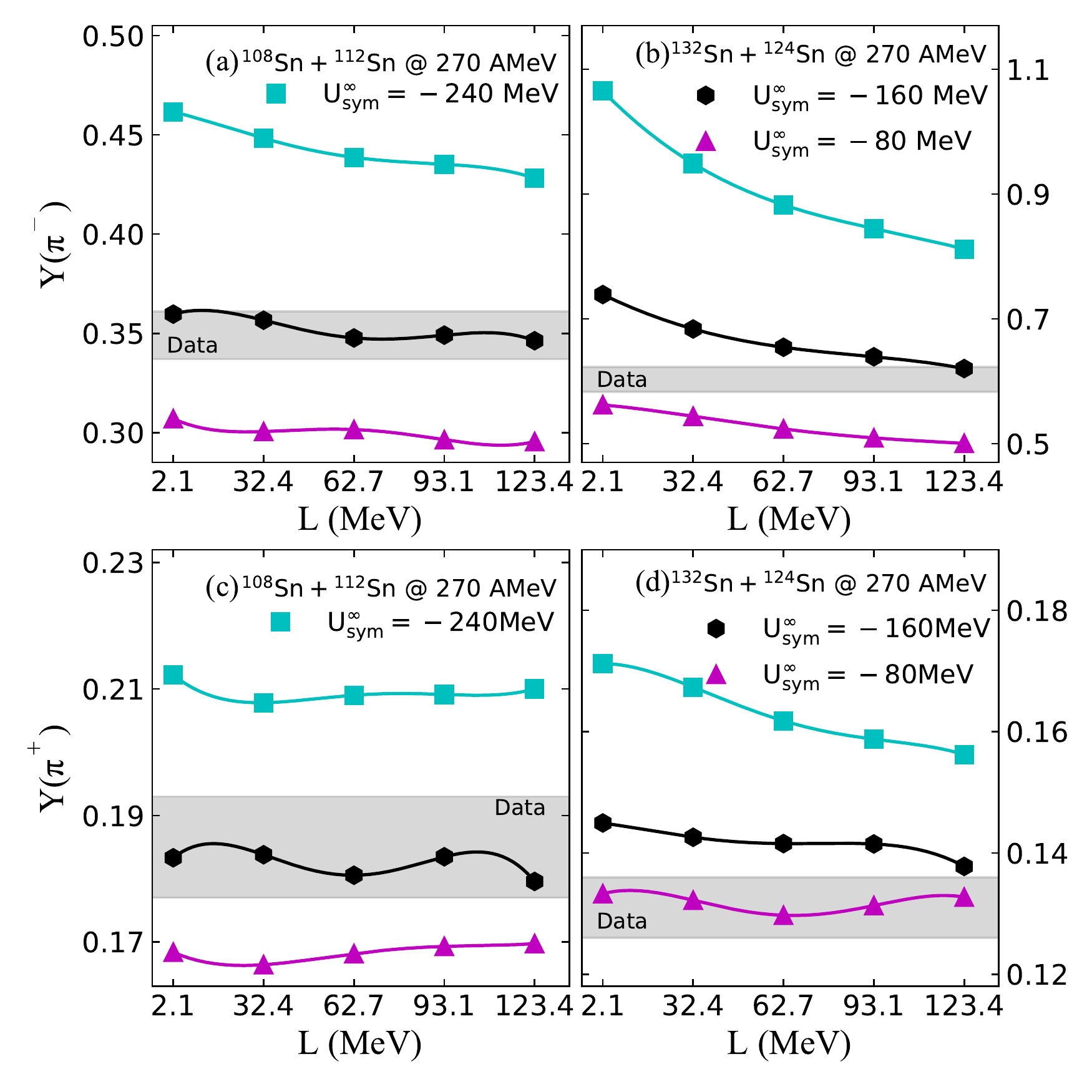}
	\caption{(Color online) Upper: Multiplicities of $\pi^{-}$ generated in reactions $^{108}$Sn + $^{112}$Sn (a) and $^{132}$Sn + $^{124}$Sn (b) with different $U_{sym}^{\infty} (\rho_{0})$ as a function of $L$ in comparison with the corresponding S$\pi$RIT data. Lower: Multiplicities of $\pi^{+}$ generated in reactions $^{108}$Sn + $^{112}$Sn (c) and $^{132}$Sn + $^{124}$Sn (d) with different $U_{sym}^{\infty} (\rho_{0})$ as a function of $L$ in comparison with the corresponding S$\pi$RIT data.
	} \label{piyields}
\end{figure}

Finally, as to the treatment of Coulomb field, we calculate the electromagnetic (EM) interactions from the Maxwell equation, i.e., ${\textbf E}=-\nabla \varphi - \partial \textbf A/\partial t$, $\textbf B=\nabla \times \textbf A$, where the scalar potential $\varphi$ and vector potential $\textbf A$ of EM fields are calculated from the resources of charges $Ze$ and currents $Ze{\textbf  v}$. 
For the detailed EM field effects in HICs, we refer readers to Refs.~\cite{Wei18a,Wei18b,Wei21} for more details.

\begin{figure}[t]
	\includegraphics[width=0.9\columnwidth]{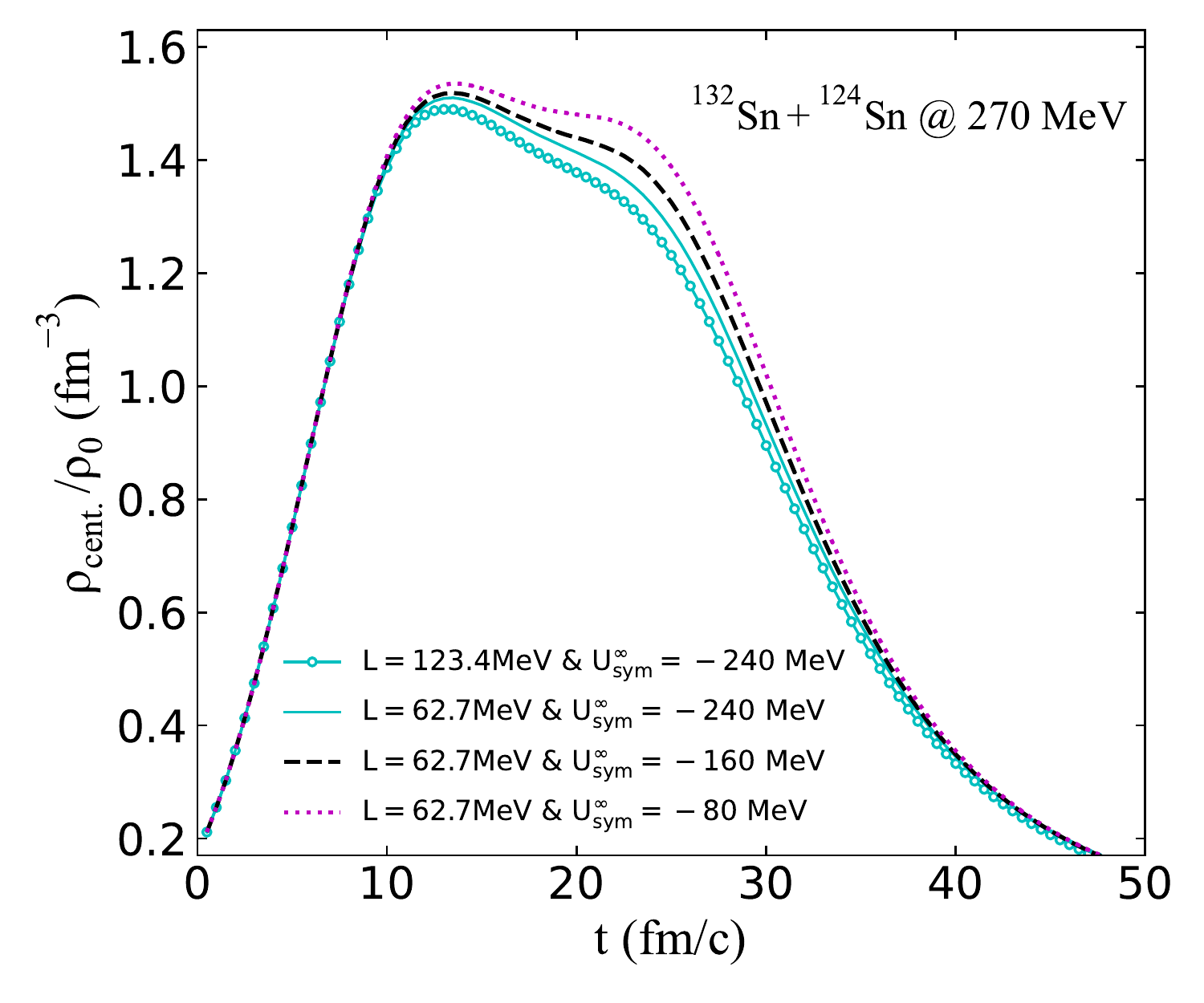}
	\caption{(Color online) Evolution of the reduced average densities in central region ($\rho_{{\rm cent.}}/\rho_{0}$) produced in $^{132}$Sn + $^{124}$Sn reactions at 270 MeV/nucleon.} \label{AveCden}
\end{figure}

\section{Results and Discussions}\label{Results and Discussions}

\begin{figure}[t]
	\includegraphics[width=0.9\columnwidth]{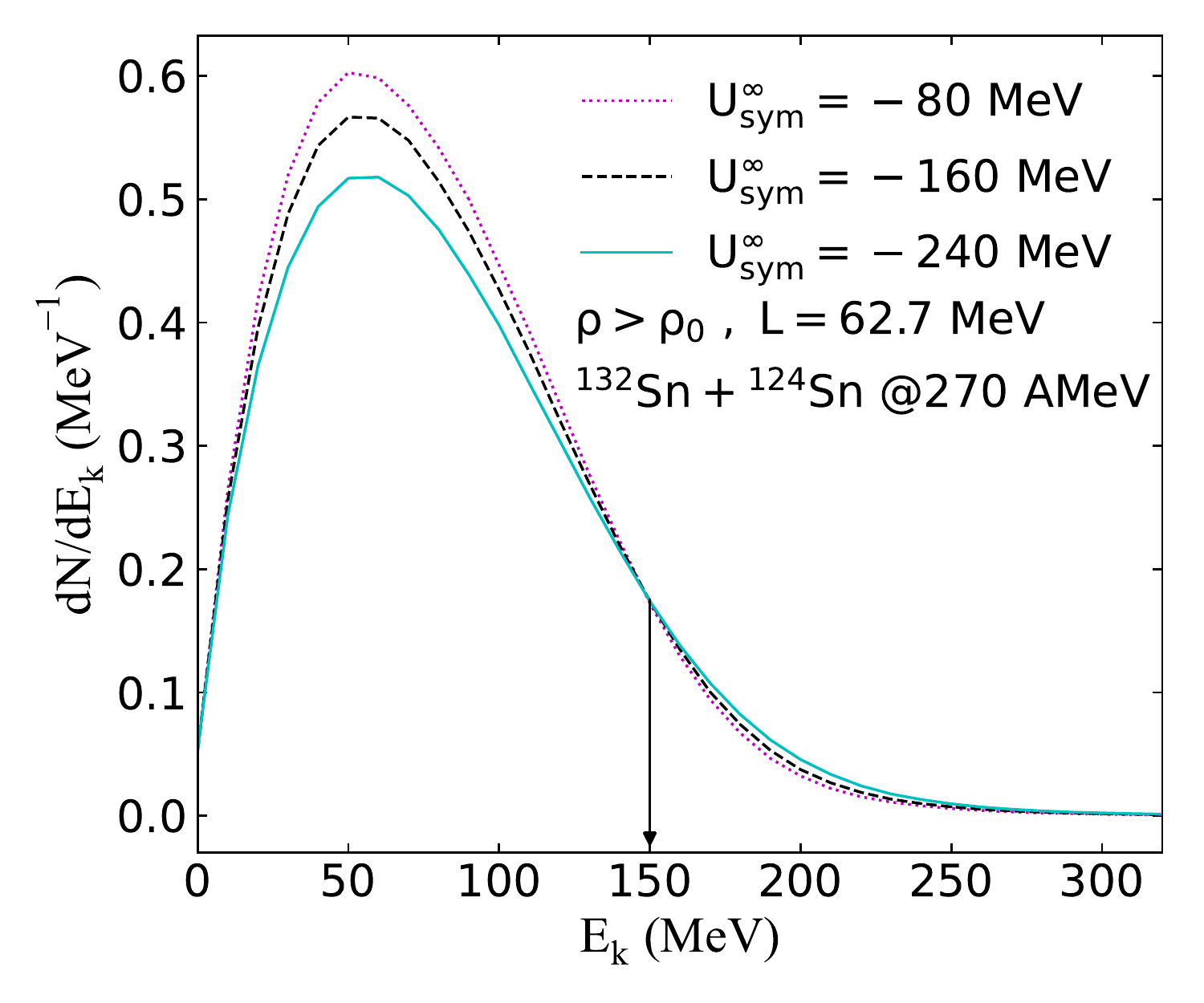}
	\caption{(Color online) Kinetic-energy distribution of nucleons in compression region at $t=20$~fm/$c$ in $^{132}$Sn + $^{124}$Sn reactions at 270 MeV/nucleon.} \label{kine}
\end{figure}

\begin{figure}[t]
	\includegraphics[width=\columnwidth]{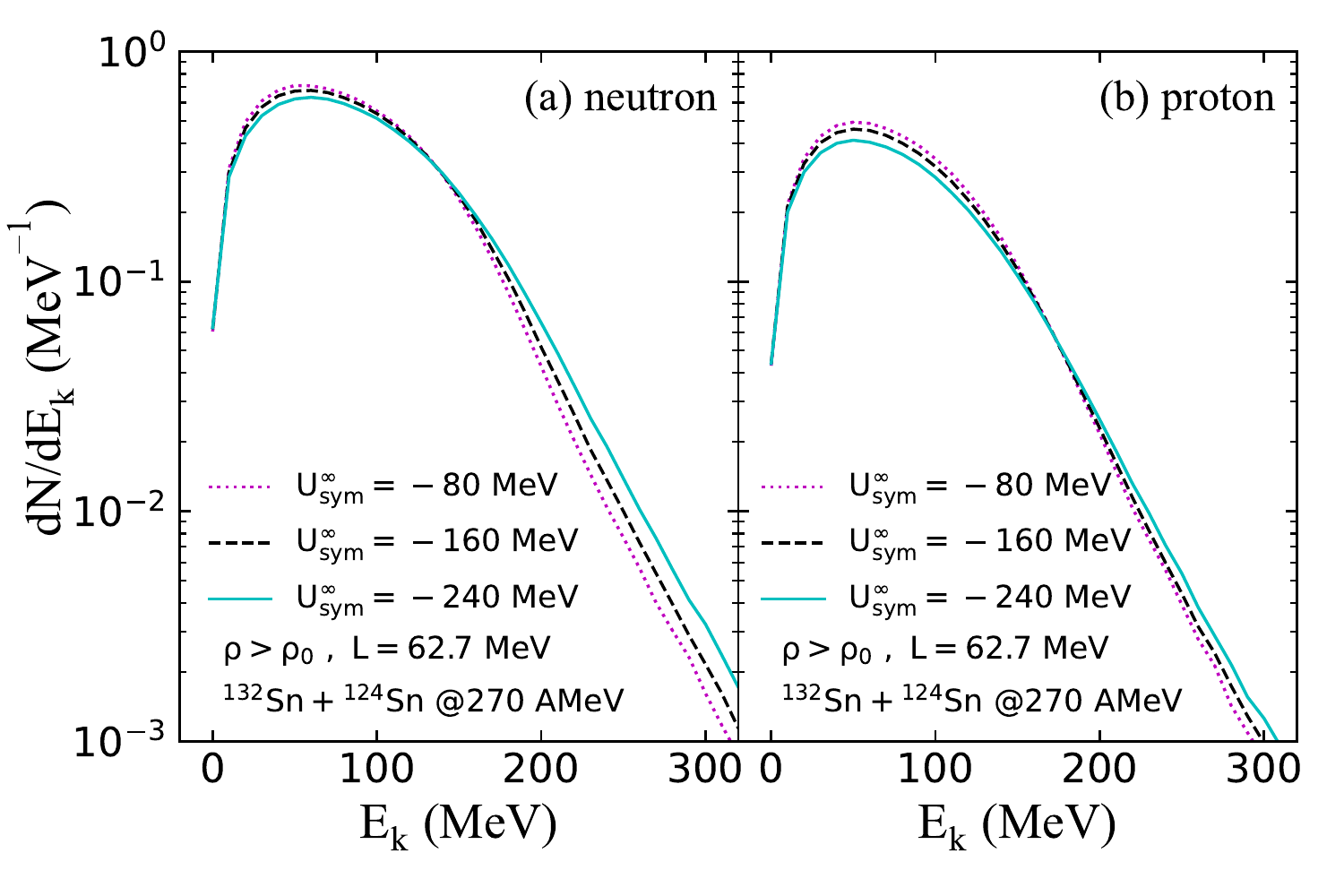}
	\caption{(Color online) Kinetic-energy distribution of neutrons (a) and protons (b) in compression region at $t=20$~fm/$c$ in $^{132}$Sn + $^{124}$Sn reactions at 270 MeV/nucleon.} \label{Ekdis}
\end{figure}

Now, we turn to the pion production in $^{108}$Sn + $^{112}$Sn and $^{132}$Sn + $^{124}$Sn reactions at 270 MeV/nucleon with an impact parameter of $b=3$ fm. 
To study the sensitivities of pion yields to the high-density behavior of \esym (i.e., $L$) and the momentum dependence of symmetry potential (i.e., $U_{sym}^{\infty}(\rho_{0})$), pion yields as a function of $L$ for different $U_{sym}^{\infty}(\rho_{0})$ are shown in Fig.~\ref{piyields}. First, consistent with the findings in Refs.~\cite{Yong21,LiBA05}, it is seen that the multiplicities of $\pi^{-}$ are more sensitive to $L$ compared to those of $\pi^{+}$, in particular for the larger isospin asymmetry reactions $^{132}$Sn + $^{124}$Sn, since $\pi^{-}$ is mostly produced from the neutron-neutron inelastic collisions~\cite{LiBA05}. Second, it is seen that with a certain $L$ the symmetry potential with larger value of $|U_{sym}^{\infty}(\rho_{0})|$ leads to more production of $\pi^{-}$ and $\pi^{+}$. To understand this observation, we first check the evolution of central region densities formed in HICs. Shown in Fig.~\ref{AveCden} are the evolutions of central reduced densities $\rho_{\rm cent.}/\rho_{0}$ formed in $^{132}$Sn + $^{124}$Sn reactions with different $U_{sym}^{\infty}(\rho_{0})$ but a certain $L$ of 62.7~MeV. For comparison, we also show the evolution of $\rho_{\rm cent.}/\rho_{0}$ for the same reaction with $L=123.4$~MeV and $U_{sym}^{\infty}(\rho_{0})=-240$~MeV. It is seen that with a certain $U_{sym}^{\infty}(\rho_{0})=-240$~MeV the soft symmetry energy with $L=62.7$~MeV leads to a higher compression compared to that with a stiff symmetry energy $L=123.4$~MeV in agreement with previous observations in many studies. Interestingly, we notice that with a certain $L=62.7$~MeV the $U_{sym}^{\infty}(\rho_{0})$ also affects the evolution of central region densities. Specifically, approximately at $13$~fm/$c$ independent of $U_{sym}^{\infty}(\rho_{0})$, the reaction with a certain $L=62.7$~MeV approaches maximum compression and thus generates a maximum compression density $1.5\rho_{0}$ in the central region, however, the decreasing velocity of this density is a little faster in case with more larger $|U_{sym}^{\infty}(\rho_{0})|$. This is due to the symmetry potential with more larger $|U_{sym}^{\infty}(\rho_{0})|$ causes some high density nucleons to gain more acceleration in the subsequent reaction stages, and thus leading to the densities of compression region to reduce slightly faster. This can be demonstrated by checking the kinetic energy distribution of nucleons in compression region with local densities higher than $\rho_{0}$ at $t=20$~fm/$c$ as shown in Fig.~\ref{kine}. Obviously, with a certain $L=62.7$~MeV but varying $U_{sym}^{\infty}(\rho_{0})$ from $-80$ to $-240$~MeV, we indeed can observe increased high energy nucleons but reduced low energy nucleons. In general, since the scalar potential has the same repulsive effects on neutrons and protons, while the symmetry potential has the repulsive (attractive) effects on the high density but low energy\footnote{Approximately at $t=13$~fm/$c$, the reaction approaches maximum compression, the nucleons in compression region are naturally in dense but low energy phase.} neutrons (protons), it is natural that one might expect these high energy nucleons to be neutrons. Nevertheless, as shown in Fig.~\ref{Ekdis}, these high energy nucleons contain both neutrons and protons, and of course, neutrons outnumber protons due to the reaction itself is neutron-rich system. Moreover, as indicated by the arrow in Fig.~\ref{kine}, the kinetic energies of these energetic neutrons and protons are above $150$~MeV, while the threshold energy of pion production through $NN$ inelastic collisions is no more than $300$~MeV. Naturally, with a certain $L$ but varying $U_{sym}^{\infty}(\rho_{0})$ from $-80$ to $-240$~MeV, we can understand the increased production of both $\pi^{-}$ and $\pi^{+}$ shown in Fig.~\ref{piyields}, since $\pi^{-}$ and $\pi^{+}$ are produced mainly from inelastic $nn\rightarrow pn\pi^{-}$ and $pp\rightarrow pn\pi^{+}$ channels. Third, compared with the S$\pi$RIT data as shown in Fig.~\ref{piyields}, our prediction on pion multiplicities with a certain range of $U_{sym}^{\infty}(\rho_{0})$ indeed can fit the experimental data on the whole. Certainly, for the predicted multiplicities, $\pi^{+}$ does not seem to be as good as $\pi^{-}$.

\begin{figure}[t]
	\includegraphics[width=\columnwidth]{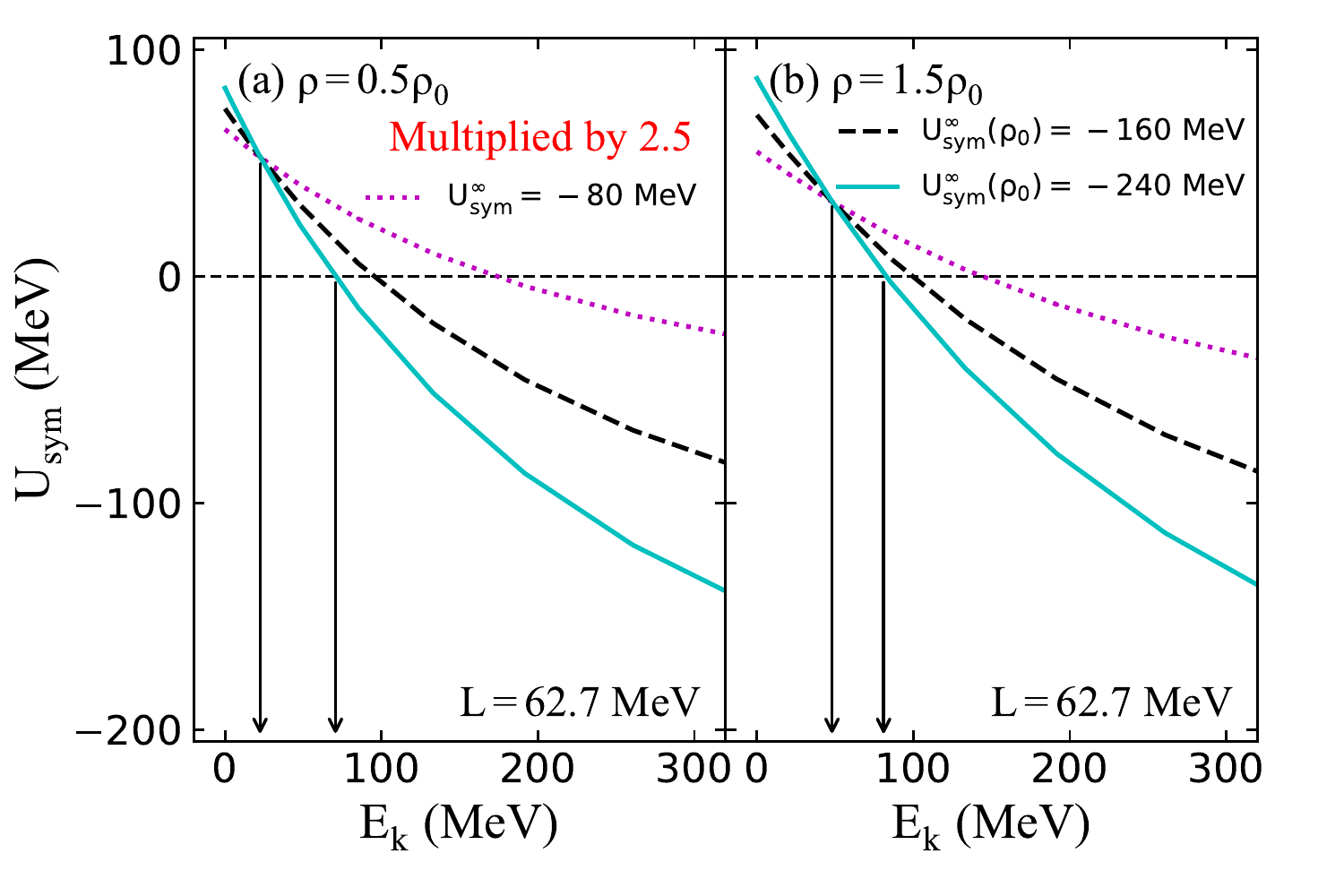}
	\caption{(Color online) Kinetic-energy dependent symmetry potentials at $\rho=0.5\rho_{0}$ (a) and $\rho=1.5\rho_{0}$ (b) with different $U_{sym}^{\infty} (\rho_{0})$ calculated from the IMDI interaction. The values of symmetry potential at $\rho=0.5\rho_{0}$ are multiplied by a factor of 2.5.} \label{sympot2}
\end{figure}

\begin{figure}[t]
	\includegraphics[width=\columnwidth]{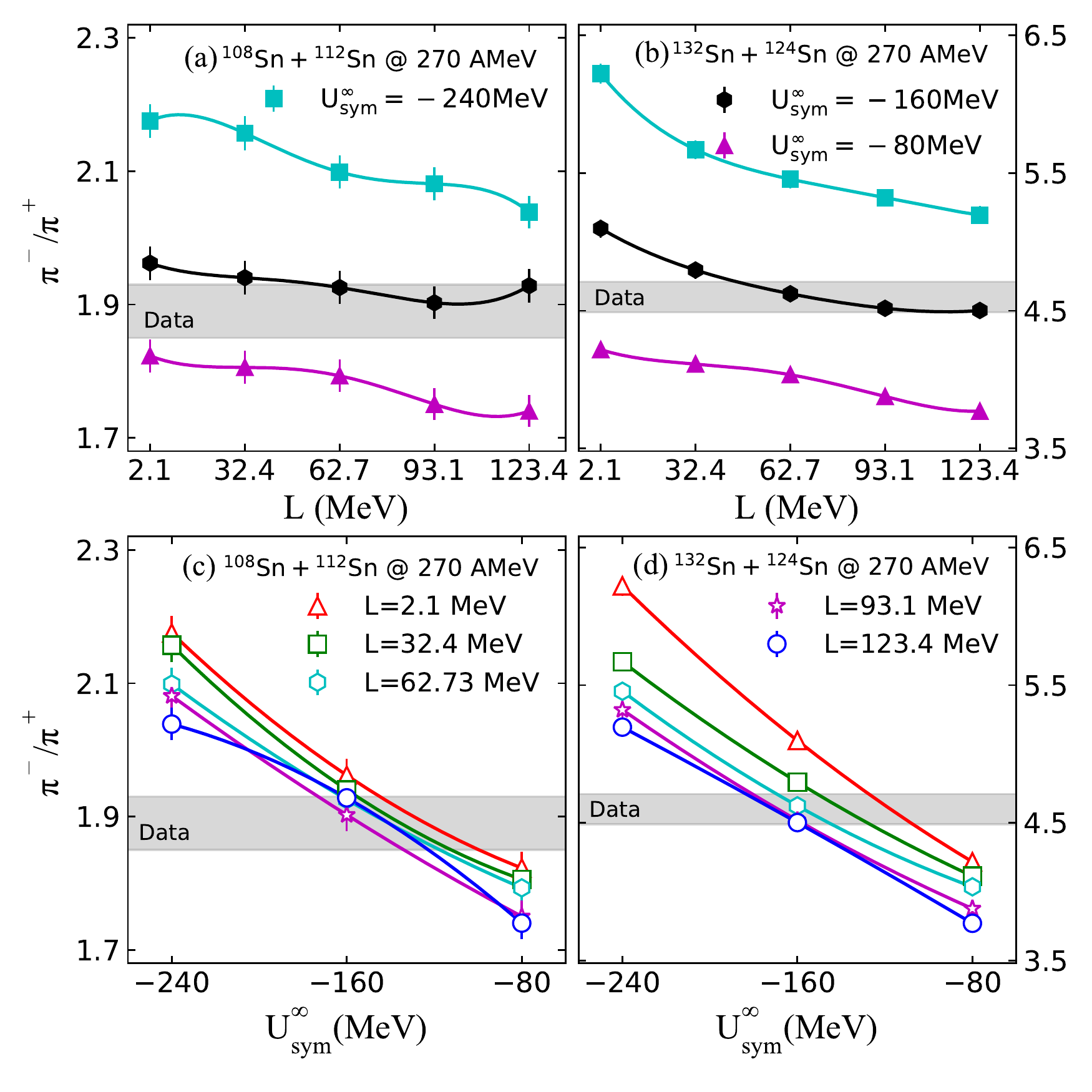}
	\caption{(Color online) Upper: Ratios of $\pi^{-}/\pi^{+}$ generated in reactions $^{108}$Sn + $^{112}$Sn (a) and $^{132}$Sn + $^{124}$Sn (b) with different $U_{sym}^{\infty} (\rho_{0})$ as a function of $L$ in comparison with the corresponding S$\pi$RIT data. Lower: Ratios of $\pi^{-}/\pi^{+}$ generated in reactions $^{108}$Sn + $^{112}$Sn (c) and $^{132}$Sn + $^{124}$Sn (d) with different $L$ as a function of $U_{sym}^{\infty} (\rho_{0})$ in comparison with the corresponding S$\pi$RIT data.} \label{ratio}
\end{figure}

So far, one might wonder the symmetry potential at $1.5\rho_{0}$ (i.e., attainable maximum densities in the compress stage) with a certain $L$ but varying $U_{sym}^{\infty}(\rho_{0})$ from $-80$ to $-240$~MeV could cause both high energy neutrons and protons to increase. In order to understand this observation, we show in right panel of Fig.~\ref{sympot2} the symmetry potential at $1.5\rho_{0}$ with a certain $L=62.7$~MeV but different $U_{sym}^{\infty}(\rho_{0})$. For completeness, we also show in left panel of Fig.~\ref{sympot2} the corresponding symmetry potential at low densities (i.e., $0.5\rho_{0}$). It is seen that similar to the symmetry potential at $\rho_{0}$, the symmetry potentials at $1.5\rho_{0}$ even with different $U_{sym}^{\infty}(\rho_{0})$ have a same value approximately at the nucleon kinetic energy of $47$~MeV. In addition, the value of symmetry potential also changes from positive to negative when the kinetic energy of nucleon is larger than a certain value depending on the value of $U_{sym}^{\infty}(\rho_{0})$. Specifically, with a certain $L=62.7$~MeV but varying $U_{sym}^{\infty}(\rho_{0})$ from $-80$ to $-240$~MeV, protons (neutrons) in high density phase will feel more stronger attractive (repulsive) effects from the symmetry potentials when their kinetic energies are lower than $47$~MeV. In contrast, if their kinetic energies are larger than $47$~MeV but lower than about $81$~MeV,\footnote{The value of $81$~MeV is the transition kinetic energy for the symmetry potential at $1.5\rho_{0}$ with $L=62.7$~MeV and $U_{sym}^{\infty}(\rho_{0})=-240$~MeV.} protons (neutrons) in high density phase will feel more {\it weaker} attractive (repulsive) effects from the symmetry potentials. Therefore, with a certain $L$ but varying $U_{sym}^{\infty}(\rho_{0})$ from $-80$ to $-240$~MeV, the repulsive scalar potential and the {\it weakened attractive} symmetry potential can cause some protons to increase their kinetic energies up to $150$~MeV and above. It should be emphasized that for the reaction with a certain $U_{sym}^{\infty}(\rho_{0})$ one usually can interpret the effects of $L$ on pion production through the density criterion, i.e., average maximum densities formed in the reaction compress stages. Nevertheless, for the case with a certain $L$, interpreting the effects of $U_{sym}^{\infty}(\rho_{0})$ on pion production needs both density and energy criterions because a small reduction of the average maximum densities formed in the reaction compress stages but a significant increase of the kinetic energy for these high density nucleons could also lead to increased production of pions.

\begin{figure}[t]
	\includegraphics[width=0.9\columnwidth]{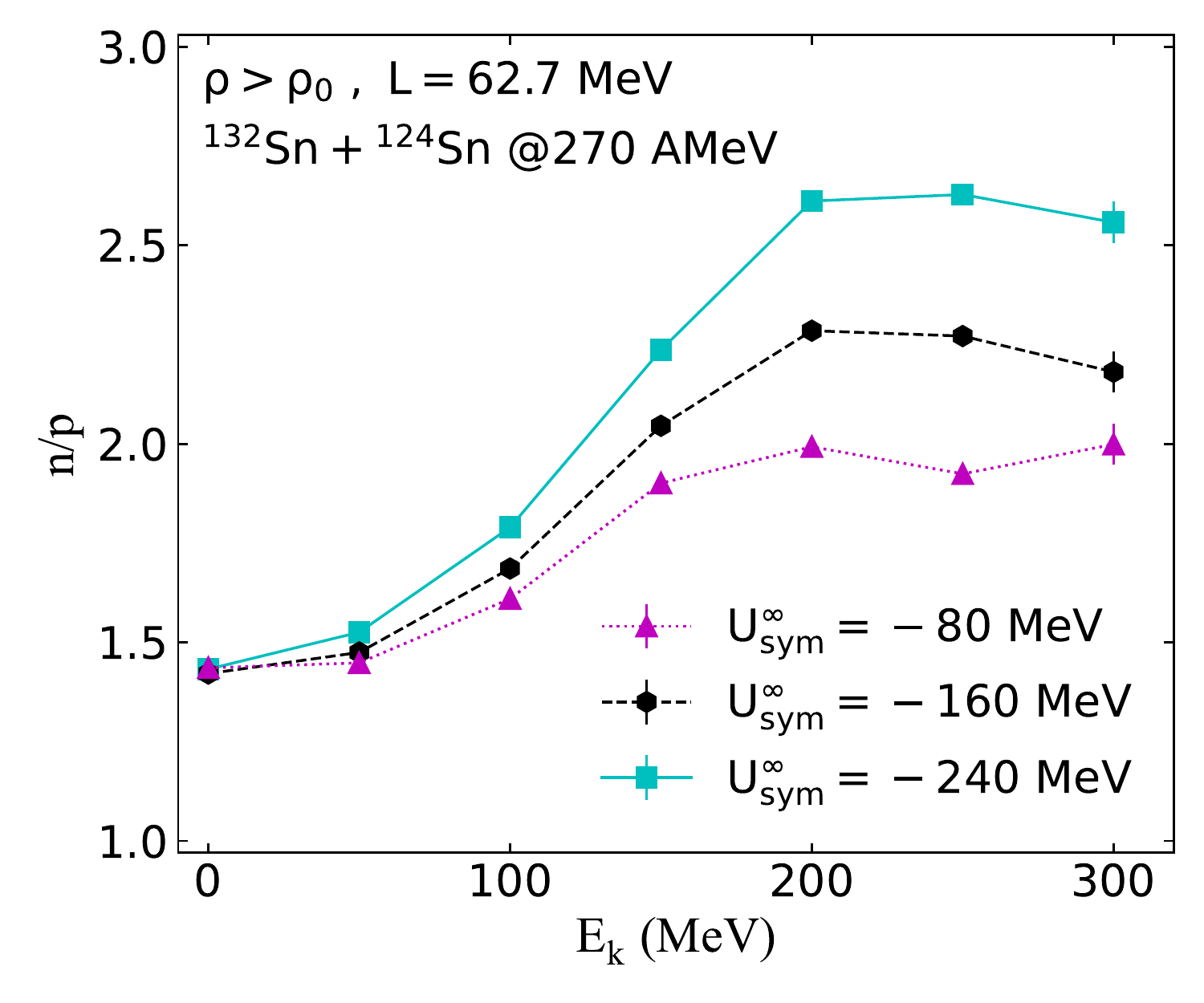}
	\caption{(Color online) Kinetic energy distribution of neutrons over protons $n/p$ with local densities higher than $\rho_{0}$ produced at $t=20$~fm/$c$ in the reaction $^{132}$Sn + $^{124}$Sn with different $U_{sym}^{\infty} (\rho_{0})$ and a certain $L$.} \label{np}
\end{figure}

Shown in Fig.~\ref{ratio} are the $\pi^{-}/\pi^{+}$ ratios of theoretical simulations for the same reactions in comparison with the S$\pi$RIT data. First, consistent with the observations of most transport models, it is seen from the upper windows of Fig.~\ref{ratio} that the $\pi^{-}/\pi^{+}$ ratios indeed are more sensitive to $L$ compared to the pion yields, and a softer symmetry energy with a smaller $L$ value leads to a higher $\pi^{-}/\pi^{+}$ ratio.
Moreover, for the more neutron-rich reaction $^{132}$Sn + $^{124}$Sn, the $\pi^{-}/\pi^{+}$ ratios show more sensitivities to $L$. Second, it is seen from the lower windows of Fig.~\ref{ratio} that with a certain $L$ the $\pi^{-}/\pi^{+}$ ratios are increasing with the value of $|U_{sym}^{\infty}(\rho_{0})|$.
Actually, similar as the reason for more pion production, this observation can also be understood by examining the kinetic energy distribution of neutrons over protons $n/p$ with local densities higher than $\rho_{0}$ at $t=20$~fm/$c$ in reactions $^{132}$Sn + $^{124}$Sn with a certain $L$ as shown in Fig.~\ref{np}. It is seen that with a certain $L=62.7$~MeV the ratio $n/p$ is increasing with varying $U_{sym}^{\infty}(\rho_{0})$ from $-80$ to $-240$~MeV, due to the increment of high energy neutrons is larger than that of protons for the neutron-rich reactions. This is the reason we can observe that with a certain $L$ the $\pi^{-}/\pi^{+}$ ratios are increasing with the value of $|U_{sym}^{\infty}(\rho_{0})|$ as shown in Fig.~\ref{ratio}. In addition, compared with the S$\pi$RIT data, our results on $\pi^{-}/\pi^{+}$ ratios also fit quite well the experimental data within a certain range for the value of $U_{sym}^{\infty}(\rho_{0})$. 

\begin{figure}[t]
	\includegraphics[width=\columnwidth]{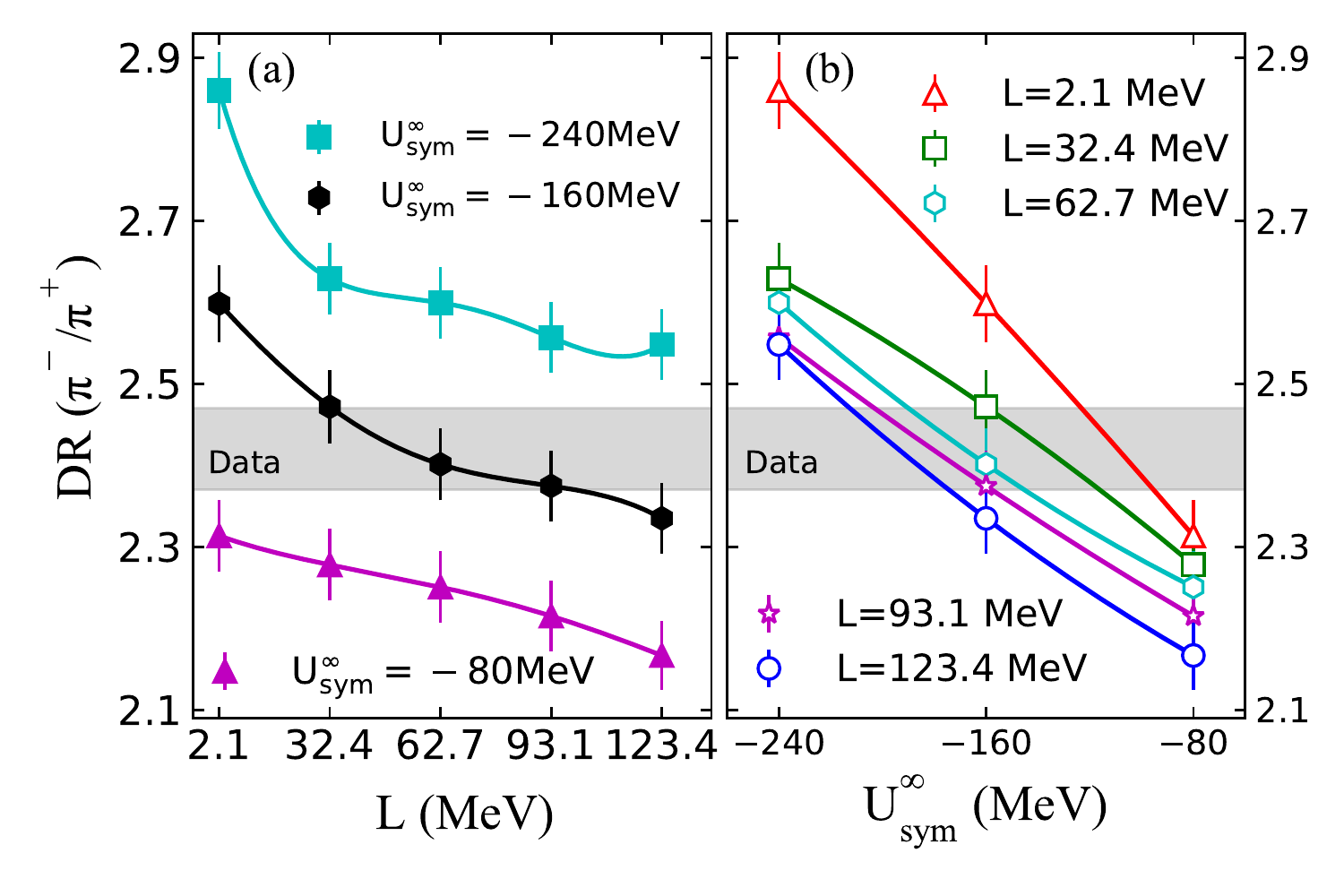}
	\caption{(Color online) The double $\pi^{-}/\pi^{+}$ ratios [i.e., DR($\pi^{-}/\pi^{+}$)] of the reactions $^{132}$Sn + $^{124}$Sn over $^{108}$Sn + $^{112}$Sn with different $U_{sym}^{\infty}(\rho_{0})$ as a function of $L$ (a) and different $L$ as a function of $U_{sym}^{\infty}(\rho_{0})$ (b) in comparison with the corresponding S$\pi$RIT data.} \label{Draito}
\end{figure}

\begin{figure}[t]
	\includegraphics[width=\columnwidth]{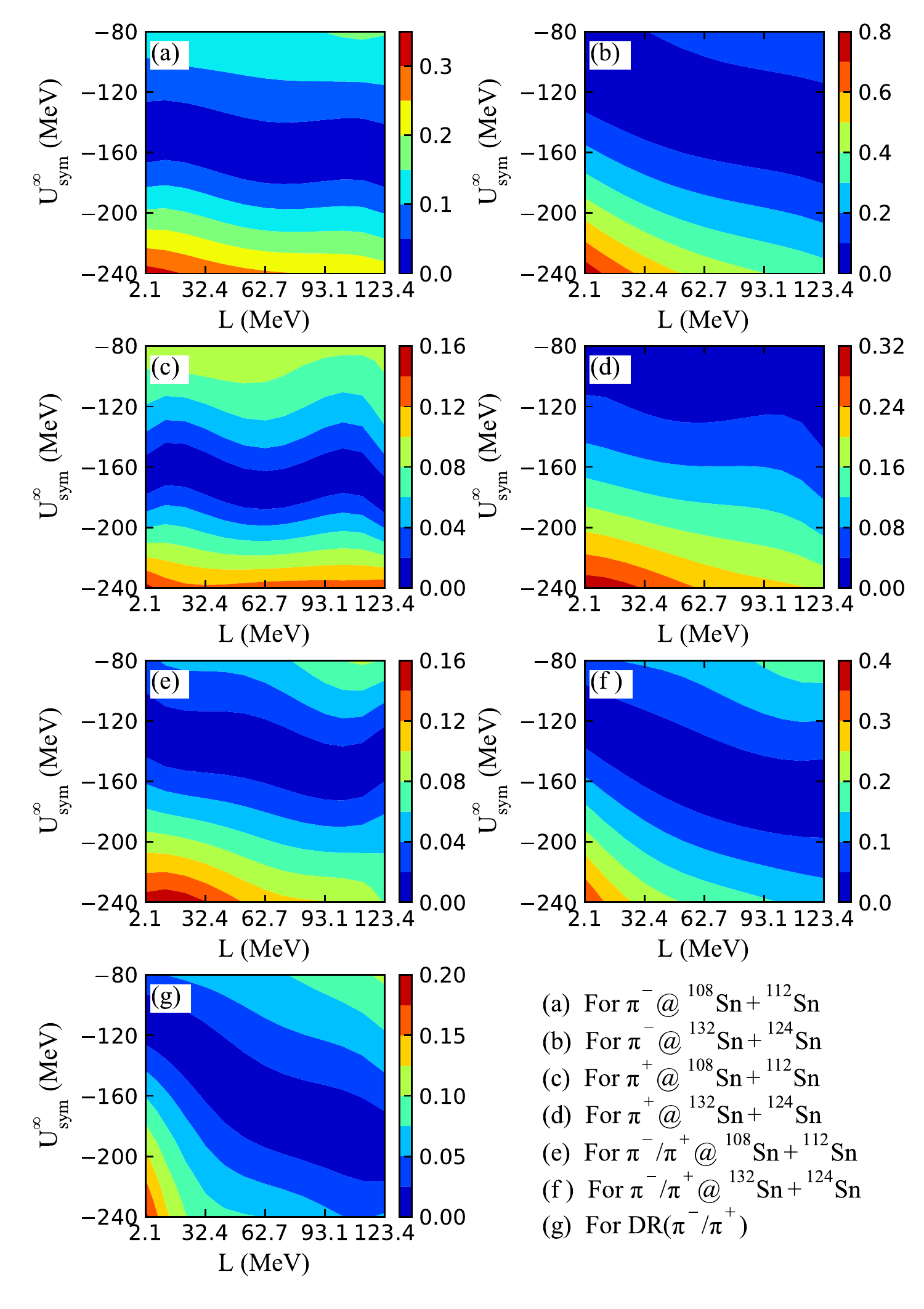}
	\caption{(Color online) Contours of the relative errors for pion yields as well as their single and double pion ratios as a function of $L$ and $U_{sym}^{\infty}(\rho_{0})$ in reactions $^{108}$Sn + $^{112}$Sn and $^{132}$Sn + $^{124}$Sn.} \label{contour}
\end{figure}

\begin{figure}[t]
	\includegraphics[width=\columnwidth]{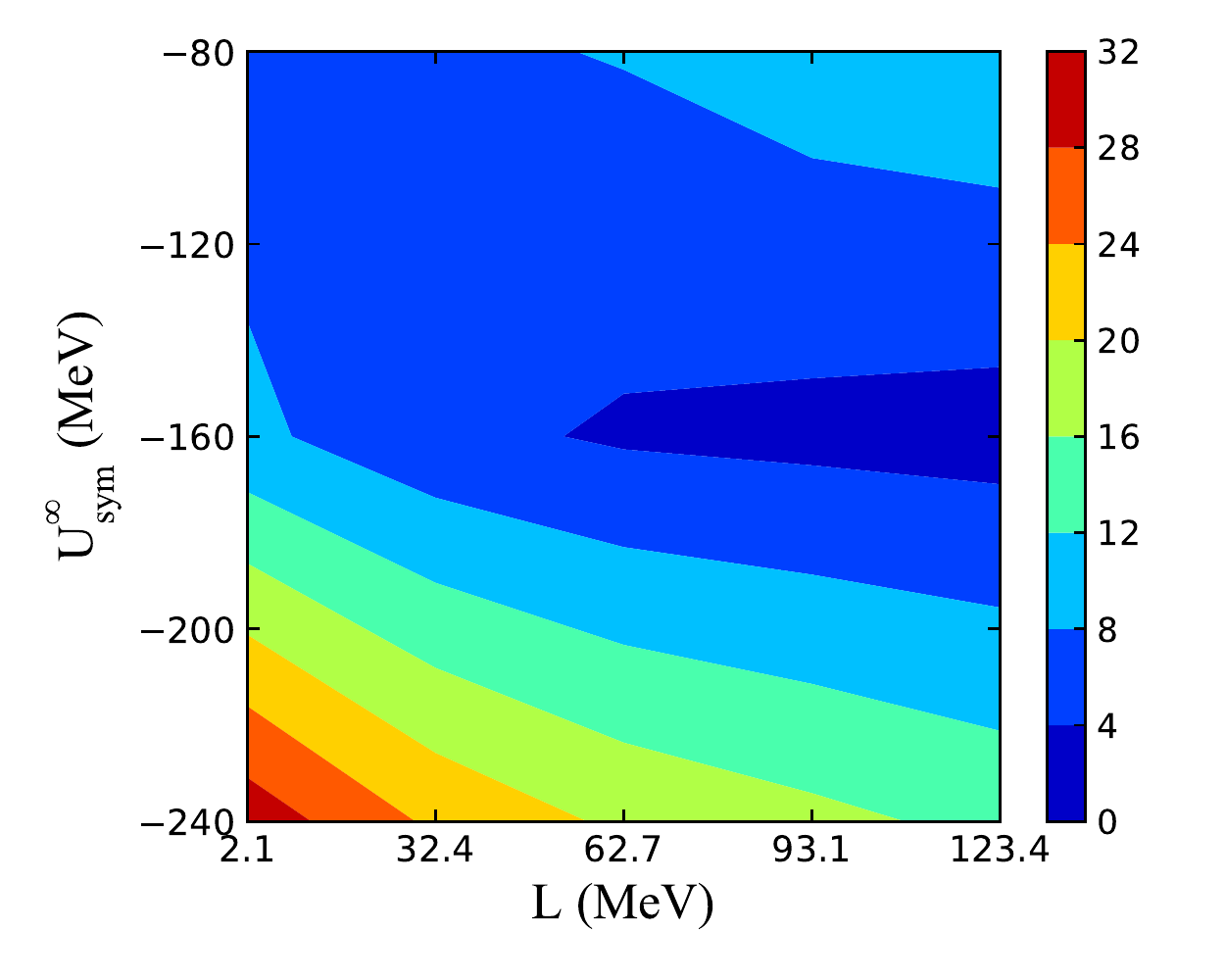}
	\caption{(Color online) The  value of $\chi$ as a two dimensional function of $U^{\infty}_{sym}(\rho_{0})$ and $L$ in reactions $^{108}$Sn + $^{112}$Sn and $^{132}$Sn + $^{124}$Sn.} \label{chi}
\end{figure}
 
\begin{figure}[t]
	\includegraphics[width=\columnwidth]{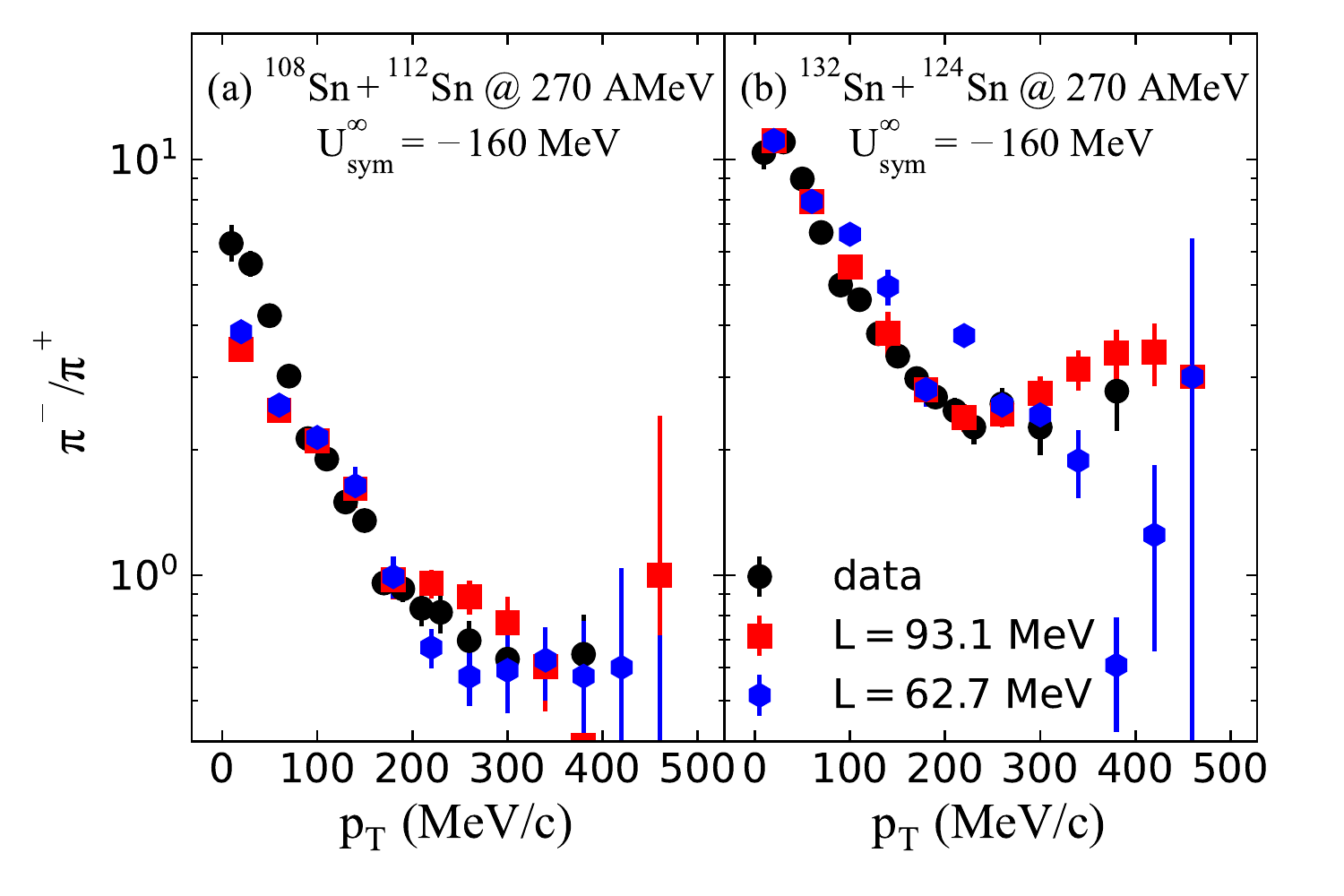}
	\caption{(Color online) The spectral pion ratios of theoretical simulations for the reactions $^{108}$Sn + $^{112}$Sn (a) and $^{132}$Sn + $^{124}$Sn (b) as a function of transverse momentum in comparison with the corresponding data.} \label{pionSpe}
\end{figure}

\begin{figure}[t]
	\includegraphics[width=\columnwidth]{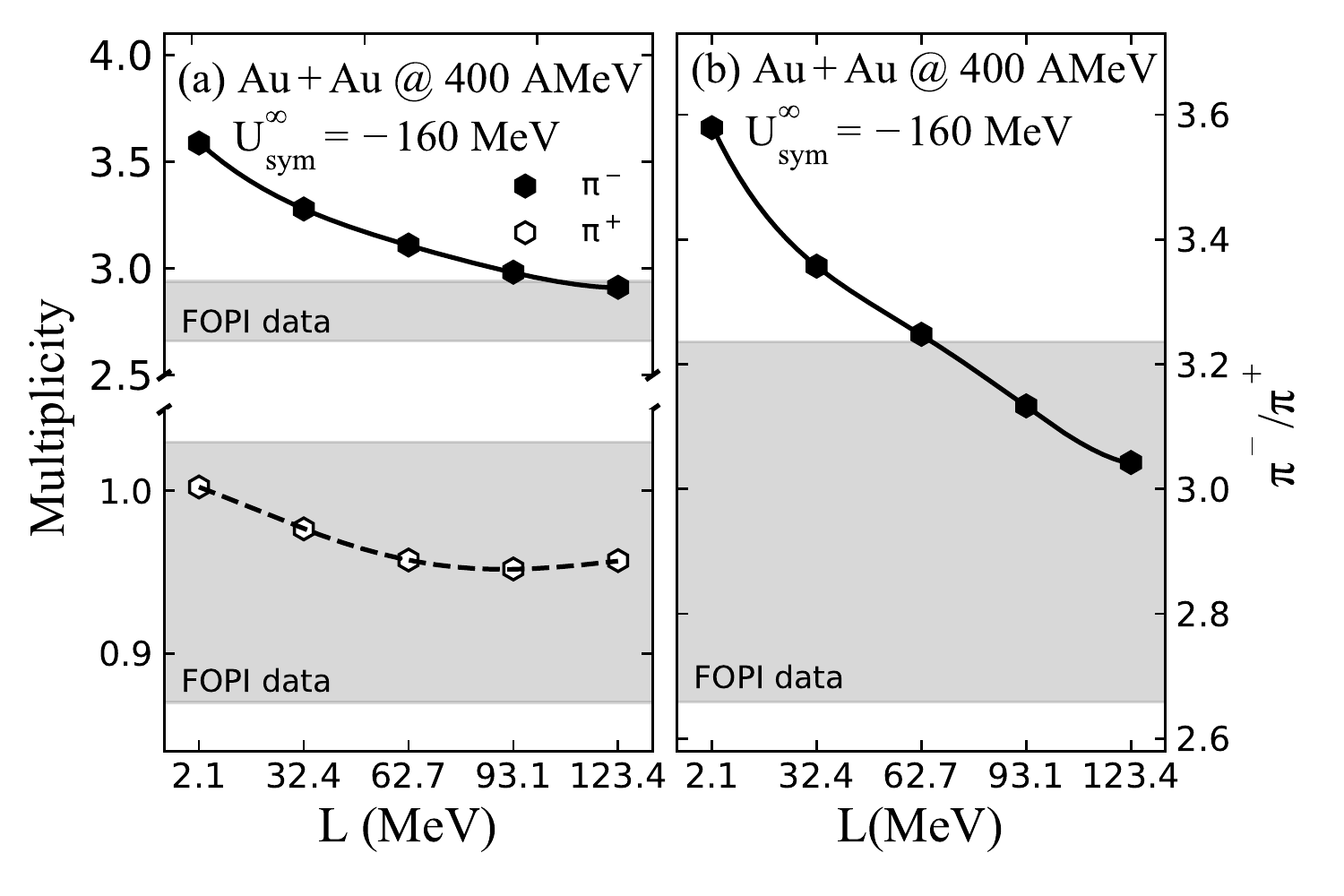}
	\caption{(Color online) Multiplicities of charged pions (a) as well their pion ratios (b) in $^{197}$Au + $^{197}$Au collisions at 400~MeV/nucleon in comparison with the corresponding data.} \label{Aupion}
\end{figure}

As a more clean observable, the double ratio of two reactions, i.e., DR($\pi^{-}/\pi^{+}$) ratio of reactions $^{132}$Sn + $^{124}$Sn over $^{108}$Sn + $^{112}$Sn, has the advantages of reducing both the isoscalar potential effects and the Coulomb field effects, and thus is expected to disentangle the effects of symmetry potential/energy from those of both isoscalar potentials and Coulomb fields in HICs. Therefore, we show in Fig.~\ref{Draito} the DR($\pi^{-}/\pi^{+}$) ratios of two reactions in comparison with the S$\pi$RIT data. It is seen from the left panel of Fig.~\ref{Draito} that the DR($\pi^{-}/\pi^{+}$) ratios of two reactions indeed are more sensitive to the high-density behavior of $E_{sym}(\rho)$. Moreover, the DR($\pi^{-}/\pi^{+}$) ratios are also more clearly separated by varying the value of $U_{sym}^{\infty}(\rho_{0})$ from -80 to -240 MeV, and thus more sensitive to the momentum dependence of symmetry potential as indicated in right panel of Fig.~\ref{Draito}. 

Now, we attempt to use above three observables, i.e., pion yields and their single $\pi^{-}/\pi^{+}$ as well as double DR($\pi^{-}/\pi^{+}$) ratios, to constrain the values of $U_{sym}^{\infty}(\rho_{0})$ and $L$. To this end, we perform the systematic error analyses for pion yields as well as their single $\pi^{-}/\pi^{+}$ and double DR($\pi^{-}/\pi^{+}$) ratios at different $U_{sym}^{\infty}(\rho_{0})$ and $L$. Apart from the pion yields as well as their single $\pi^{-}/\pi^{+}$ and double DR($\pi^{-}/\pi^{+}$) ratios at $U_{sym}^{\infty}(\rho_{0})=-80, -160$ and $-240$ MeV and $L=2.1$, $32.4$, $62.7$, $93.1$ and $123.4$~MeV, the values of these observables at other $U_{sym}^{\infty}(\rho_{0})$ and $L$ with an interval of 10 MeV are obtained by interpolating the simulation ones.  Shown in Fig.~\ref{contour} are contours of the relative error as a two dimensional function of $L$ and $U_{sym}^{\infty}(\rho_{0})$. Unfortunately, it seems hard to constrain the values of $L$ and $U_{sym}^{\infty}(\rho_{0})$ simultaneously from the relative errors between theoretical simulations and experimental data. Therefore, we further perform systematic $\chi-$square analyses for these observables at different $L$ and $U_{sym}^{\infty}(\rho_{0})$, the corresponding $\chi$ values as a two dimensional  function of $L$ and $U_{sym}^{\infty}(\rho_{0})$ are shown in Fig.~\ref{chi}. It is seen that there is an area with highest confidence, in which $U_{sym}^{\infty}(\rho_{0})$ is constrained to be $-160^{+18}_{-9}$~MeV, and the lower limit of $L$ is approximately no less than 55~MeV.


To further verify the above results, we take a value of $-160$~MeV for $U_{sym}^{\infty}(\rho_{0})$ to check the spectral pion ratios of the reactions $^{108}$Sn + $^{112}$Sn and $^{132}$Sn + $^{124}$Sn, since this observable especially its high-energy parts might be the best probe of high-density behavior of \esym as shown in Refs.~\cite{Estee21,trans5}. Shown in Fig.~\ref{pionSpe} are the spectral pion ratios of theoretical simulations in comparison with the corresponding data~\cite{Estee21}. It is seen that with this value for $U_{sym}^{\infty}(\rho_{0})$ the spectral pion ratios of our simulations, especially its high-energy parts, indeed can fit fairly the experimental data when $L$ ranging from 62.7 to 93.1~MeV. In addition, we can also see that the pion observable in $^{197}$Au + $^{197}$Au collisions at 400~MeV/nucleon also supports this value for $U_{sym}^{\infty}(\rho_{0})$ as shown in Fig.~\ref{Aupion}. On the other hand, it is well known that the isospin splitting of in-medium nucleon effective mass is resulting from the momentum dependence of symmetry potential. Therefore, it is useful to evaluate the isospin splitting of in-medium nucleon effective mass for our used $U_{sym}^{\infty}(\rho_{0})$. According to the formula of nucleon effective mass, i.e., 
\begin{equation}\label{effective-mass}
m^{*}_{\tau}/m=\Big{[}1+\frac{m}{k_{\tau}}\frac{dU_{\tau}}{dk}\Big{]}^{-1},
\end{equation}
for the used $U_{sym}^{\infty}(\rho_{0})=-10$, $-80$, $-160$ and $-240$~MeV, the corresponding neutron-proton effective mass splittings $\Delta m_{np}^{*}$ are 0.178$\delta$, 0.384$\delta$, 0.622$\delta$, and 0.864$\delta$, respectively.

So far, one can find that our results suggest a constraint on $L$, i.e., $62.7<L<93.1$~MeV. It can be seen that this constraint on $L$ is very close to the values $70<L<101$~MeV extracted from charge exchange and elastic scattering reactions in Ref.~\cite{Dan17}, and also within the constrained range of $42<L<117$~MeV in Ref.~\cite{Estee21}. Moreover, our results also suggest a constraint on $U_{sym}^{\infty}(\rho_{0})$, i.e., $-160^{+18}_{-9}$~MeV. Certainly, we also notice that our extracted value for $U_{sym}^{\infty}(\rho_{0})$, e.g., $-160$~MeV, leads to a larger isospin splitting than the upper limit 0.33$\delta$ in Refs.~\cite{Estee21,Xu17}. The reasons might be two folds. First, authors of Ref.~\cite{Estee21} use a more accurate criterion, i.e., $E_{sym}(2\rho_{0}/3)=25.5$~MeV~\cite{Cozma18}, and thus consider the uncertainties of $E_{sym}(\rho_{0})$, i.e.,  $32.5<E_{sym}(\rho_{0})<38.1$~MeV; while in this study, we use a fixed value $32.5$~MeV for $E_{sym}(\rho_{0})$ as commonly used, and thus do not consider its uncertainties.
Second, since the separate density-dependent scenario for in-medium nucleon-nucleon interaction has been used in this study as aforementioned, the corresponding potential energy density and single nucleon potential in Eqs.~(\ref{IMDI}) and (\ref{IMDIU}) as well the corresponding expressions of $A_{l}$ and $A_{u}$ are different from those in Ref.~\cite{Xu17}. The two aspects might lead to the difference between our extracted isospin splitting and that in Refs.~\cite{Estee21,Xu17}. As indicated in Refs.~\cite{Zhang18,Stone17}, the accurate inclusion of these effects might be important for transport model simulations, and thus may further improve our results, e.g., the deviations of simulated $\pi^{+}$ from the corresponding data.
Finally, it should be mentioned that our results in the present work are mainly based on the S$\pi$RIT experiments. It will also be interesting to see how the $U_{sym}^{\infty}(\rho_{0})$ affects observables measured in other experiments such as FOPI experiments~\cite{FOPI} as well ASY-EOS experiments\cite{Russ16}. 

\section{Summary}\label{Summary}

In conclusion, we have studied effects of the momentum dependence of symmetry potential on pion production in central Sn + Sn collisions at 270 MeV/nucleon. It is found that with a certain $L$ the characteristic parameter $U_{sym}^{\infty}(\rho_{0})$ of momentum dependent symmetry potential affects significantly the production of $\pi^{-}$ and $\pi^{+}$ as well as their pion ratios. Moreover, through performing systematic analyses of these observables as well comparing the spectral pion ratios of theoretical simulations with the experimental data, we find a constraint on $L$, i.e., $62.7<L<93.1$~MeV. Also, the $U_{sym}^{\infty}(\rho_{0})$ is constrained to be $-160^{+18}_{-9}$~MeV.
In addition, it is shown that the pion observable of $^{197}$Au + $^{197}$Au collisions at 400~MeV/nucleon also supports the extracted value for $U_{sym}^{\infty}(\rho_{0})$. 

\begin{acknowledgments}
G.F.W. would like to thank Profs. B. A. Li and G. C. Yong for helpful discussions.
This work is supported by the National Natural Science Foundation of China under grant Nos.11965008, 11405128, and Guizhou Provincial Science and Technology Foundation under Grant No.[2020]1Y034, and the PhD-funded project of Guizhou Normal university (Grant No.GZNUD[2018]11).
\end{acknowledgments}


\begin{thebibliography}{99}
	
\bibitem{Typel01} S. Typel, B. A. Brown, Neutron radii and the neutron equation of state in relativistic models. Phys. Rev. C \textbf{64}, 027302 (2001). \url{https://doi.org/10.1103/PhysRevC.64.027302}

\bibitem{kolo05}E. E. Kolomeitsev, C. Hartnack, H. W. Barz
{\it et al}., 
Transport theories for heavy-ion collisions in the 1 A GeV regime. J. Phys. G:Nucl. Part. Phys. \textbf{31}, S741 (2005). \url{https://doi.org/10.1088/0954-3899/31/6/015}

\bibitem{ditoro} V. Baran, M. Colonna, V. Greco
{\it et al}., Reaction dynamics with exotic nuclei. Phys. Rep. \textbf{410}, 335 (2005). \url{https://doi.org/10.1016/j.physrep.2004.12.004}

\bibitem{LCK08} B.A. Li, L.W. Chen, C.M. Ko, Recent progress and new challenges in isospin physics with heavy-ion reactions. Phys. Rep. \textbf{464}, 113 (2008). \url{https://doi.org/10.1016/j.physrep.2008.04.005}

\bibitem{Tam11} A. Tamii, I. Poltoratska, P. von-Neumann-Cosel
{\it et al}., Complete electric dipole response and the neutron skin in 
$^{208}$Pb. Phys. Rev. Lett. \textbf{107}, 062502 (2011). \url{https://doi.org/10.1103/PhysRevLett.107.062502}

\bibitem{Vin14}X. Vi\~{n}as, M. Centelles, X. Roca-Maza
{\it et al}., Density dependence of the symmetry energy from neutron skin thickness in finite nuclei. Eur. Phys. J. A \textbf{50}, 27 (2014). \url{https://doi.org/10.1140/epja/i2014-14027-8}

\bibitem{Hor14} C. J. Horowitz, E. F. Brown, Y. Kim
{\it et al}., A way forward in the study of the symmetry energy: experiment, theory, and observation. J. Phys. G:Nucl. Part. Phys. \textbf{41}, 093001 (2014). \url{https://doi.org/10.1088/0954-3899/41/9/093001}

\bibitem{Rein16} P. G. Reinhard, W. Nazarewicz, Nuclear charge and neutron radii and nuclear matter: Trend analysis in Skyrme density-functional-theory approach. Phys. Rev. C \textbf{93}, 051303 (2016). \url{https://doi.org/10.1103/PhysRevC.93.051303}

\bibitem{Baldo16} M. Baldo, G. F. Burgio, The nuclear symmetry energy. Prog. Part. Nucl. Phys. \textbf{91}, 203 (2016). \url{https://doi.org/10.1016/j.ppnp.2016.06.006}

\bibitem{MCW18}C. W. Ma, Y. G. Ma, Shannon information entropy in heavy-ion collisions. Prog. Part. Nucl. Phys. \textbf{99}, 120 (2018). \url{https://doi.org/10.1016/j.ppnp.2018.01.002}

\bibitem{Yu20}H. Yu, D. Q. Fang, Y. G. Ma, Investigation of the symmetry energy of nuclear matter using isospin-dependent quantum molecular dynamics. Nucl. Sci. Tech. 31, 61 (2020). \url{https://doi.org/10.1007/s41365-020-00766-x}

\bibitem{MCW21}C. W. Ma, H. L. Wei, X. Q. Liu
{\it et al}., Nuclear fragments in projectile fragmentation reactions. Prog. Part. Nucl. Phys. \textbf{121}, 103911 (2021). \url{https://doi.org/10.1016/j.ppnp.2021.103911}

\bibitem{Estee21} J. Estee, W. G. Lynch, C. Y. Tsang
{\it et al}., Probing the Symmetry Energy with the Spectral Pion Ratio. Phys. Rev. Lett. \textbf{126}, 162701 (2021). \url{https://doi.org/10.1103/PhysRevLett.126.162701}

\bibitem{Tsang19} C. Y. Tsang, M. B. Tsang, P. Danielewicz
{\it etal}., Insights on Skyrme parameters from GW170817. Phys. Lett. B \textbf{796}, 1 (2019). \url{https://doi.org/10.1016/j.physletb.2019.05.055}

\bibitem{Lim18} Y. Lim, J. W. Holt, Neutron star tidal deformabilities constrained by nuclear theory and experiment. Phys. Rev. Lett. \textbf{121}, 062701 (2018). \url{https://doi.org/10.1103/PhysRevLett.121.062701}

\bibitem{Tews18} I. Tews, J. Margueron, S. Reddy, Critical examination of constraints on the equation of state of dense matter obtained from GW170817. Phys. Rev. C \textbf{98}, 045804 (2018). \url{https://doi.org/10.1103/PhysRevC.98.045804}

\bibitem{Drago14} A. Drago, A. Lavagno, G. Pagliara
{\it et al}., Early appearance of $\Delta$ isobars in neutron stars. Phys. Rev. C \textbf{90}, 065809 (2014). \url{https://doi.org/10.1103/PhysRevC.90.065809}

\bibitem{Steiner12} A. W. Steiner, S. Gandolfi, Connecting neutron star observations to three-body forces in neutron matter and to the nuclear symmetry energy. Phys. Rev. Lett. \textbf{108}, 081102 (2012). \url{https://doi.org/10.1103/PhysRevLett.108.081102}

\bibitem{Duco11} C. Ducoin, J. Margueron, C. Providência
{\it et al}., Core-crust transition in neutron stars: Predictivity of density developments. Phys. Rev. C \textbf{83}, 045810 (2011). \url{https://doi.org/10.1103/PhysRevC.83.045810}

\bibitem{Latt16} J. M. Lattimer, M. Prakash, The equation of state of hot, dense matter and neutron stars. Phys. Rep. \textbf{621}, 127 (2016). \url{https://doi.org/10.1016/j.physrep.2015.12.005}

\bibitem{Brown13}B. A. Brown, Constraints on the Skyrme equations of state from properties of doubly magic nuclei. Phys. Rev. Lett. \textbf{111}, 232502 (2013). \url{https://doi.org/10.1103/PhysRevLett.111.232502}

\bibitem{LiBA16}B. A. Li, B. J. Cai, L. W. Chen, {\it et al}., Isospin dependence of nucleon effective masses in neutron-rich matter. Nucl. Sci. Tech. 27, 141 (2016). \url{https://doi.org/10.1007/s41365-016-0140-4}

\bibitem{Dan02} P. Danielewicz, R. Lacey, W. G. Lynch, Determination of the equation of state of dense matter. Science \textbf{298}, 1592 (2002). \url{https://doi.org/10.1126/science.1078070}

\bibitem{Oert17} M. Oertel, M. Hempel, T. KI\"{a}hn
{\it et al}, Equations of state for supernovae and compact stars. Rev. Mod. Phys. \textbf{89}, 015007 (2017). \url{https://doi.org/10.1103/RevModPhys.89.015007}

\bibitem{Cai17}B. J. Cai, L. W. Chen, Constraints on the skewness coefficient of symmetric nuclear matter within the nonlinear relativistic mean field model. Nucl. Sci. Tech. 28, 185 (2017). \url{https://doi.org/10.1007/s41365-017-0329-1}

\bibitem{Wei20a}G. F. Wei, Q. J. Zhi, X. W. Cao
{\it et al}., Examination of an isospin-dependent single-nucleon momentum distribution for isospin-asymmetric nuclear matter in heavy-ion collisions. Nucl. Sci. Tech. 31, 71 (2020). \url{https://doi.org/10.1007/s41365-020-00779-6}

\bibitem{Liu21}J. Liu, C. Gao, N. Wan
{\it et al}., Basic quantities of the equation of state in isospin asymmetric nuclear matter. Nucl. Sci. Tech. 32, 117 (2021). \url{https://doi.org/10.1007/s41365-021-00955-2}

\bibitem{Hoff72} G. W. Hoffmann, W. R. Coker, Coupled-Channel Calculations of the Energy Dependence of the ($p$,$n$) Charge-Exchange Reaction. Phys. Rev. Lett. \textbf{29}, 227 (1972). \url{https://doi.org/10.1103/PhysRevLett.29.227}

\bibitem{Kon03} A. J. Koning, J. P. Delaroche, Local and global nucleon optical models from 1 KeV to 200 MeV. Nucl. Phys. A \textbf{713}, 231 (2003). \url{https://doi.org/10.1016/S0375-9474(02)01321-0}

\bibitem{Jeu91} J. P. Jeukenne, C. Mahaux, R. Sartor, Dependence of the Fermi energy upon neutron excess. Phys. Rev. C \textbf{43}, 2211 (1991). \url{https://doi.org/10.1103/PhysRevC.43.2211}

\bibitem{Jhang21}G. Jhang, J. Estee, J. Barney
{\it et al}., Symmetry energy investigation with pion production from Sn+Sn systems. Phys. Lett. \textbf{B} 813, 136016 (2021). \url{https://doi.org/10.1016/j.physletb.2020.136016}

\bibitem{Shane15}R. Shane, A. B. McIntosh, T. Isobe,
{\it et al}., 
S$\pi$RIT: A time-projection chamber for symmetry-energy studies. 
Nucl. Instr. Meth. A \textbf{784}, 513 (2015). \url{https://doi.org/10.1016/j.nima.2015.01.026}

\bibitem{FOPI} W. Reisdorf, A. Andronic, R. Averbeck {\it et al.}, Systematics of central heavy ion collisions in the 
regime. 
Nucl. Phys. A \textbf{848}, 366 (2010). \url{https://doi.org/10.1016/j.nuclphysa.2010.09.008}

\bibitem{Yong21} G. C. Yong, Symmetry energy extracted from the S$\pi$RIT pion data in Sn + Sn systems. Phys. Rev. C \textbf{104}, 014613 (2021). \url{https://doi.org/10.1103/PhysRevC.104.014613}

\bibitem{Subedi08}R. Subedi, R. Shneor, P. Monaghan
{\it et al.}, Probing cold dense nuclear matter. Science \textbf{320}, 1476 (2008). \url{https://doi.org/10.1126/science.1156675}

\bibitem{Wein11}L. B. Weinstein, E. Piasetzky, D. W. Higinbotham
{\it et al}., Short range correlations and the EMC effect. Phys. Rev. Lett. \textbf{106}, 052301 (2011). \url{https://doi.org/10.1103/PhysRevLett.106.052301}

\bibitem{Sar14}M. M. Sargsian, New properties of the high-momentum distribution of nucleons in asymmetric nuclei. Phys. Rev. C \textbf{89}, 034305 (2014). \url{https://doi.org/10.1103/PhysRevC.89.034305}

\bibitem{Ciofi15}C. Ciofi degli Atti, In-medium short-range dynamics of nucleons: Recent theoretical and experimental advances. Phys. Rep. \textbf{590}, 1 (2015). \url{https://doi.org/10.1016/j.physrep.2015.06.002}

\bibitem{Ohen14}O. Hen, M. Sargsian, L. B. Weinstein
{\it et al.}, Momentum sharing in imbalanced Fermi systems. Science \textbf{346}, 614 (2014). \url{https://doi.org/10.1126/science.1256785}

\bibitem{Ohen18}M. Duer, O. Hen, E. Piasetzky
{\it et al.}, Probing high-momentum protons and neutrons in neutron-rich nuclei. Nature \textbf{560}, 617 (2018). \url{https://doi.org/10.1038/s41586-018-0400-z}

\bibitem{Brue64}K. A. Brueckner, J. Dabrowski, Symmetry energy and the isotopic spin dependence of the single-particle potential in nuclear matter. Phys. Rev. \textbf{134}, B722 (1964). \url{https://doi.org/10.1103/PhysRev.134.B722}

\bibitem{Dabr73}J. Dabrowski, P. Haensel, Spin and spin-isospin symmetry energy of nuclear matter. Phys. Rev. C \textbf{7}, 916 (1973). \url{https://doi.org/10.1103/PhysRevC.7.916}


\bibitem{Gior10}V. Giordano, M. Colonna, M. D. Toro
{\it et al}., Isospin emission and flow at high baryon density: A test of the symmetry potential. Phys. Rev. C \textbf{81}, 044611 (2010). \url{https://doi.org/10.1103/PhysRevC.81.044611}

\bibitem{trans1}J. Xu, L. W. Chen, M. B. Tsang
{\it et al}., Understanding transport simulations of heavy-ion collisions at 100 A and 400 A MeV: Comparison of heavy-ion transport codes under controlled conditions. Phys. Rev. C \textbf{93}, 044609 (2016). \url{https://doi.org/10.1103/PhysRevC.93.044609}

\bibitem{trans2}Y. X. Zhang, Y. J. Wang, M. Colonna
{\it et al}., Comparison of heavy-ion transport simulations: Collision integral in a box. Phys. Rev. C \textbf{97}, 034625 (2018). \url{https://doi.org/10.1103/PhysRevC.97.034625}

\bibitem{trans3}A. Ono, J. Xu, M. Colonna
{\it et al}., Comparison of heavy-ion transport simulations: Collision integral with pions and $\Delta$ resonances in a box. Phys. Rev. C \textbf{100}, 044617 (2019). \url{https://doi.org/10.1103/PhysRevC.100.044617}

\bibitem{trans4}M. Colonna, Y. X. Zhang, Y. J. Wang
{\it et al}., Comparison of heavy-ion transport simulations: Mean-field dynamics in a box. Phys. Rev. C \textbf{104}, 024603 (2021). \url{https://doi.org/10.1103/PhysRevC.104.024603}

\bibitem{trans5}H. Wolter, M. Colonna, D. Cozma
{\it et al}., Transport model comparison studies of intermediate-energy heavy-ion collisions. Prog. Part. Nucl. Phys. \textbf{125}, 103962 (2022). \url{https://doi.org/10.1016/j.ppnp.2022.103962}

\bibitem{Das03} C. B. Das, S. Das Gupta, C. Gale
{\it et al.}, Momentum dependence of symmetry potential in asymmetric nuclear matter for transport model calculations. Phys. Rev. C \textbf{67}, 034611 (2003). \url{https://doi.org/10.1103/PhysRevC.67.034611}

\bibitem{IBUU}B. A. Li, C. B. Das, S. Das Gupta
{\it et al.}, Momentum dependence of the symmetry potential and nuclear reactions induced by neutron-rich nuclei at RIA. Phys. Rev. C \textbf{69}, 011603(R) (2004). \url{https://doi.org/10.1103/PhysRevC.69.011603}

\bibitem{CLnote} L. W. Chen, B. A. Li, A note of an improved MDI interaction for transport model simulations of heavy ion collisions (Unpublished, Texas A\&M University-Commerce, 2010).

\bibitem{Xu10}C. Xu, B. A. Li, Improved single particle potential for transport model simulations of nuclear reactions induced by rare isotope beams. Phys. Rev. C \textbf{81}, 044603 (2010). \url{https://doi.org/10.1103/PhysRevC.81.044603}

\bibitem{Chen14}L. W. Chen, C. M. Ko, B. A. Li
{\it et al.}, Probing isospin- and momentum-dependent nuclear effective interactions in neutron-rich matter. Eur. Phys. J. A \textbf{50}, 29 (2014). \url{https://doi.org/10.1140/epja/i2014-14029-6}

\bibitem{Wei20} G. F. Wei, C. Xu, W. Xie
{\it et al.}, Effects of density-dependent scenarios of in-medium nucleon-nucleon interactions in heavy-ion collisions. Phys. Rev. C \textbf{102}, 024614 (2020). \url{https://doi.org/10.1103/PhysRevC.102.024614}

\bibitem{Gogny80}J. Decharg\'{e}, D. Gogny, Hartree-Fock-Bogolyubov calculations with the D1 effective interaction on spherical nuclei. Phys. Rev. C \textbf{21}, 1568 (1980). \url{https://doi.org/10.1103/PhysRevC.21.1568}

\bibitem{Duguet03}T. Duguet, P. Bonche, Density dependence of two-body interactions for beyond–mean-field calculations. Phys. Rev. C \textbf{67}, 054308 (2003). \url{https://doi.org/10.1103/PhysRevC.67.054308}

\bibitem{Negele70}J. W. Negele, Structure of finite nuclei in the local-density approximation. Phys. Rev. C \textbf{1}, 1260 (1970). \url{https://doi.org/10.1103/PhysRevC.1.1260}

\bibitem{Xu15} J. Xu, L. W. Chen, B. A. Li, Thermal properties of asymmetric nuclear matter with an improved isospin- and momentum-dependent interaction. Phys. Rev. C \textbf{91}, 014611 (2015). \url{https://doi.org/10.1103/PhysRevC.91.014611}

\bibitem{Xu15b}H. Y. Kong, Y. Xia, J. Xu
{\it et al.}, Reexamination of the neutron-to-proton-ratio puzzle in intermediate-energy heavy-ion collisions. Phys. Rev. C \textbf{91}, 047601 (2015). \url{https://doi.org/10.1103/PhysRevC.91.047601}

\bibitem{Xu17} H. Y. Kong, J. Xu, L. W. Chen
{\it et al.}, Constraining simultaneously nuclear symmetry energy and neutron-proton effective mass splitting with nucleus giant resonances using a dynamical approach. Phys. Rev. C \textbf{95}, 034324 (2017). \url{https://doi.org/10.1103/PhysRevC.95.034324}

\bibitem{Hama90} S. Hama, B. C. Clark, E. D. Cooper
{\it et al.}, Global Dirac optical potentials for elastic proton scattering from heavy nuclei. Phys. Rev. C \textbf{41}, 2737 (1990). \url{https://doi.org/10.1103/PhysRevC.41.2737}

\bibitem{Buss12} O. Buss, T. Gaitanos, K. Gallmeister
{\it et al.}, Transport-theoretical description of nuclear reactions. Phys. Rep. \textbf{512}, 1 (2012). \url{https://doi.org/10.1016/j.physrep.2011.12.001}

\bibitem{Eric66}M. Ericson, T. E. O. Ericson, Optical properties of low-energy pions in nuclei. Ann. of Phys. \textbf{36}, 323 (1966). \url{https://doi.org/10.1016/0003-4916(66)90302-2}

\bibitem{Oset88}C. Garc\'{\i}a-Recio, E. Oset, L. L. Salcedo, S-wave optical potential in pionic atoms. Phys. Rev. C \textbf{37}, 194 (1988). \url{https://doi.org/10.1103/PhysRevC.37.194}

\bibitem{Oset93} J. Nieves, E. Oset, C. Garc\'{\i}a-Recio, Many-body approach to low-energy pion-nucleus scattering. Nucl. Phys. A \textbf{554}, 554 (1993). \url{https://doi.org/10.1016/0375-9474(93)90246-T}

\bibitem{Zhang17} Z. Zhang, C. M. Ko, Medium effects on pion production in heavy ion collisions. Phys. Rev. C \textbf{95}, 064604 (2017). \url{https://doi.org/10.1103/PhysRevC.95.064604}

\bibitem{Wei21} G. F. Wei, C. Liu, X. W. Cao
{\it et al.}, Necessity of self-consistent calculations for the electromagnetic field in probing the nuclear symmetry energy using pion observables in heavy-ion collisions. Phys. Rev. C \textbf{103}, 054607 (2021). \url{https://doi.org/10.1103/PhysRevC.103.054607}

\bibitem{Wei18a}G. F. Wei, B. A. Li, G. C. Yong
{\it et al.}, Effects of retarded electrical fields on observables sensitive to the high-density behavior of the nuclear symmetry energy in heavy-ion collisions at intermediate energies. Phys. Rev. C \textbf{97}, 034620 (2018). \url{https://doi.org/10.1103/PhysRevC.97.034620}

\bibitem{Wei18b}G. F. Wei, G. C. Yong, L. Ou
{\it et al.}, Beam-energy dependence of the relativistic retardation effects of electrical fields on the $\pi^{-}/\pi^{+}$ ratio in heavy-ion collisions. Phys. Rev. C \textbf{98}, 024618 (2018). \url{https://doi.org/10.1103/PhysRevC.98.024618}

\bibitem{LiBA05}B. A. Li, G. C. Yong, W. Zuo, Near-threshold pion production with radioactive beams. Phys. Rev. C \textbf{71}, 014608 (2005). \url{https://doi.org/10.1103/PhysRevC.71.014608}


\bibitem{Dan17} P. Danielewicz, P. Singh, J. Lee, Symmetry energy III: Isovector skins. Nucl. Phys. A \textbf{958}, 147 (2017). \url{https://doi.org/10.1016/j.nuclphysa.2016.11.008}

\bibitem{Cozma18}M. D. Cozma, Feasibility of constraining the curvature parameter of the symmetry energy using elliptic flow data. Eur. Phys. J. A \textbf{54}, 40 (2018). \url{https://doi.org/10.1140/epja/i2018-12470-1}

\bibitem{Zhang18}Z. Zhang, C. M. Ko, Effects of energy conservation on equilibrium properties of hot asymmetric nuclear matter. Phys. Rev. C \textbf{97}, 014610 (2018). \url{https://doi.org/10.1103/PhysRevC.97.014610}

\bibitem{Stone17}J. R. Stone, P. Danielewicz, Y. Iwata, Proton and neutron density distributions at supranormal density in low- and medium-energy heavy-ion collisions. Phys. Rev. C \textbf{96}, 014612 (2017). \url{https://doi.org/10.1103/PhysRevC.96.014612}

\bibitem{Russ16} P. Russotto, S. Gannon, S. Kupny
{\it et al.}, Results of the ASY-EOS experiment at GSI: The symmetry energy at suprasaturation density. Phys. Rev. C \textbf{94}, 034608 (2016). \url{https://doi.org/10.1103/PhysRevC.94.034608}




	
\end{thebibliography}
\end{document}